\documentclass[a4paper,11pt]{article}
\pdfoutput=1
\usepackage{graphicx}  
\usepackage{dcolumn}   
\usepackage{bm,relsize}        
\usepackage{amssymb, amsmath}
\usepackage{wasysym}
\usepackage{slashed}
\usepackage{lipsum, color}
\usepackage{jheppub} 

\usepackage[usenames,dvipsnames,svgnames,table]{xcolor}

\def\lsim{\mathrel{\rlap{\lower3pt\hbox{\hskip0pt$\sim$}}
   \raise1pt\hbox{$<$}}}         
\def\gsim{\mathrel{\rlap{\lower4pt\hbox{\hskip1pt$\sim$}}
   \raise1pt\hbox{$>$}}}         

\newcommand{\bea}{\begin{eqnarray}}
\newcommand{\eea}{\end{eqnarray}}
\newcommand{\be}{\begin{equation}}
\newcommand{\ee}{\end{equation}}
\newcommand{\bit}{\begin{itemize}}
\newcommand{\eit}{\end{itemize}}

\def\eeq{\end{equation}}

\title{
The second Higgs at the lifetime frontier
}
\author[a,b]{Samuel~Alipour-fard,}
\affiliation[a]{Department of Physics, University of California, Santa Barbara, CA 93106, USA}
\affiliation[b]{Center for Theoretical Physics, Massachusetts Institute of Technology, Cambridge, MA 02139, USA}
\author[a]{Nathaniel~Craig,}
\author[c]{Stefania~Gori,}
\affiliation[c]{Santa Cruz Institute for Particle Physics, University of California, Santa Cruz, CA 95064, USA}
\author[a]{Seth~Koren,} 
\author[d,e]{Diego Redigolo}
\affiliation[d]{Raymond and Beverly Sackler School of Physics and Astronomy, Tel-Aviv University, Tel-Aviv
69978, Israel}
\affiliation[e]{School of Natural Sciences, Institute for Advanced Study, Einstein Drive,
Princeton, NJ 08540, USA}
\date{\today}
\def\beq{\begin{equation}}
\def\eeq{\end{equation}}
\begin{document}

\abstract{
We assess the current coverage and the future discovery potential of LHC searches for heavy Higgs bosons decaying into long-lived particles (LLPs), focusing primarily on the production of pairs of LLPs with hadronic final states. These signatures are generic in dark sectors where a heavy scalar decays into pairs of lighter states which subsequently mix with the Standard Model Higgs. We show that a handful of existing analyses provide broad coverage of LLP decay lengths ranging from millimeters to tens of meters, and explore the complementarity between searches for displaced and prompt final states in several simplified models. For both heavy singlet and heavy doublet scalars, LLP searches typically provide the leading sensitivity in current data and exhibit the strongest discovery potential in future LHC runs. We further translate the impact of these searches into the parameter space of various Twin Higgs models, demonstrating that LLP searches are a promising avenue for discovering a Twin Higgs with displaced decays. Finally, we propose a variety of additional search channels that would improve coverage of the second Higgs at the lifetime frontier.  }

\maketitle

\section{Introduction}

With the discovery of a Standard Model-like Higgs in 2012 \cite{Aad:2012tfa, Chatrchyan:2012xdj}, the search for additional Higgs bosons has become a key component of the physics program at the Large Hadron Collider (LHC). The existence of such additional Higgs states is strongly motivated by many approaches to physics beyond the Standard Model, including proposed solutions to the electroweak hierarchy problem such as supersymmetry. Even apart from specific ultraviolet motivations, the diversity of spin-half and spin-one states observed in the Standard Model naturally suggests searching for similar diversity in the spin-zero sector. 

There is now an extensive suite of LHC searches for additional Higgs bosons, covering essentially all conventional decays of additional Higgs bosons to promising Standard Model final states (including the 125 GeV Higgs itself). However, considerably less attention has been devoted to potential {\it exotic} decays of additional Higgs bosons, in which the final states are quite unlike those expected from the direct couplings of new Higgs states to Standard Model particles. This provides a natural avenue for the further development of searches for additional Higgs bosons, a necessary step for ensuring complete experimental coverage of extended Higgs sectors at the LHC.

Despite their novelty, such exotic decay modes are far from exceptional in motivated extensions of the Higgs sector.\footnote{See e.g.~\cite{Chang:2005ht, Han:2007ae, Batell:2011pz, Holdom:2014boa} for select examples and previous studies. 
Of course, exotic decay modes are also far from exceptional in the Standard Model Higgs sector itself. For that matter, see \cite{Curtin:2013fra} for a survey of possible exotic decay modes of the 125 GeV Higgs.} Exotic decays can arise within an extended Higgs sector itself, due to the presence of multiple scalars with diverse masses and couplings \cite{Coleppa:2014hxa, Kling:2016opi, Coleppa:2017lue}. More broadly, very often the same frameworks that extend the Higgs sector also introduce new degrees of freedom charged under the Standard Model, which may then appear in Higgs decays. Not only do these new states give rise to exotic decays of heavy Higgs bosons, but the exotic decays may provide the main discovery channels for both the heavy Higgses and the new states alike. For example, supersymmetric extensions of the Standard Model predict a whole host of partner particles of Standard Model fields, some of which (most notably the electroweakinos) may be predominantly produced by decays of heavy supersymmetric Higgs bosons (for recent studies see e.g. \cite{Craig:2015jba,Barman:2016kgt,Kulkarni:2017xtf,Bahl:2018zmf,Gori:2018pmk}). Diverse approaches to dark matter, baryogenesis, and the hierarchy problem entail rich hidden sectors coupled primarily through the Higgs portal to additional Higgs-like scalars. Far from being an exclusive property of isolated models, such exotic decay modes are a generic feature of theories that extend more than just the Higgs sector of the Standard Model.

The lifetime of new states produced in exotic decays of additional Higgs bosons can range from prompt to stable, giving rise to a range of experimental signatures. Among the most distinctive signatures are those of long-lived particles (LLPs), whose decays within the detector volume set them qualitatively apart from promptly-decaying or detector-stable particles.\footnote{Such Hidden Valley signatures \cite{Strassler:2006im} arise readily in an extended Higgs sector \cite{Chang:2005ht, Strassler:2006ri, Han:2007ae}. For a recent summary of theory motivation for LLPs, see \cite{Curtin:2018mvb}. For a summary of past, present, and future experimental searches for LLPs, see \cite{Lee:2018pag}. For interpretations and forecasts of LHC searches for 125 GeV Higgs decays to LLPs, see \cite{Clarke:2015ala, Csaki:2015fba, Curtin:2015fna, Pierce:2017taw}.} The exotic decay products of additional Higgs bosons may be rendered long-lived by any of the properties that lead to long-lived particles in the Standard Model, such as small mass splittings, small decay couplings, off-shell decays, approximate symmetries, or combinations thereof. Once produced, the signatures of LLPs are often sufficiently distinctive that they may be identified by analyses with little or no Standard Model backgrounds, making them a promising channel for discovering additional Higgs bosons. Likewise, additional Higgses provide a promising production mode for LLPs. While decays of the 125 GeV Higgs to LLPs may be difficult to discover due to trigger thresholds (a notable limitation of LHC searches for Higgs decays into LLPs at $\sqrt{s} = 8$ TeV), decays of additional heavy Higgs bosons can be much more spectacular. 

In pursuing a search program for heavy Higgs bosons decaying into LLPs, a natural consideration is the complementarity between Higgs decays into LLPs and direct decays into Standard Model final states. Appreciable production of heavy Higgses typically implies couplings that induce direct Standard Model decay modes, and such couplings can arise either intrinsically (if additional Higgs multiplets carry Standard Model quantum numbers) or through mixing with the Standard Model-like Higgs (if they are Standard Model singlets). A robust program of searches for heavy Higgses in conventional final states is sensitive to these direct decays, making it valuable to develop a framework in which the reach in prompt and displaced final states can be compared.

Perhaps the two simplest frameworks for exploring heavy Higgs decays into LLPs are extensions of the Higgs sector by a singlet scalar or an additional electroweak doublet scalar, respectively. In each case, the additional scalar can couple in turn to pairs of long-lived particles. The signatures of the singlet scalar model 
are determined by a relatively small number of free parameters, namely the mass of the additional physical Higgs scalar, the mixing between the singlet scalar and the Standard Model-like Higgs, the strength of the singlet coupling to $hh$ relative to $WW/ZZ$, and the strength of the coupling to LLPs. While there are more free parameters in the scalar doublet model, these may be further reduced by assuming one Higgs doublet couples to all Standard Model fermions while the other couples to the LLPs. An additional motivated simplification is to work in the so-called alignment limit \cite{Gunion:2002zf,Delgado:2013zfa,Craig:2013hca,Carena:2013ooa,Haber:2013mia} in which the couplings of the 125 GeV Higgs are exactly Standard Model-like; this guarantees consistency with Standard Model Higgs coupling measurements and dictates that the additional CP-even Higgs scalar couples only to Standard Model fermions. In this case the free parameters are the masses of the additional physical Higgs scalars, the strength of the couplings to LLPs, and the parameter $\tan \beta$ controlling the distribution of vacuum expectation values between the doublets. As for the LLPs themselves, while they possess a variety of possible decay modes, for singlet scalar LLPs the most natural option is for them to decay via mixing with the SM-like Higgs. In this case they inherit the branching ratios of a Standard Model Higgs of the same mass, leaving the LLP mass and proper lifetime as free parameters. Taken together, these considerations define a pair of simple frameworks for jointly interpreting bounds on heavy Higgs decays into LLPs, bounds on heavy Higgs decays to Standard Model states, and constraints coming from coupling measurements of the 125 GeV Higgs.

These simplified models can be mapped on to diverse extensions of the Standard Model that involve heavy Higgs decays to LLPs, including proposed solutions to the electroweak hierarchy problem. In supersymmetric theories, for example, the lightest electroweakino can become a long-lived particle in the presence of $R$-parity violation, giving rise to LLP production in heavy supersymmetric Higgs decays. Heavy Higgs decays to LLPs play an even more prominent role in theories of neutral naturalness \cite{Craig:2014aea} such as the Twin Higgs \cite{Chacko:2005pe}, where the additional Higgs bosons required to stabilize the weak scale generically decay into hidden sector bound states that travel macroscopic distances before returning to the Standard Model. As we will see, searches for heavy Higgs decays into LLPs may provide some of the strongest constraints on (and greatest discovery potential for) these theories. 

In this work we initiate a systematic study of a second Higgs boson at the lifetime frontier. In Section \ref{sec:modelInd}, we develop a model-independent parameterization of a second scalar resonance and use it to illustrate the impact of existing LLP searches at the LHC and their complementarity with searches for prompt decays of a heavy Higgs. We then construct two simple models to facilitate the interpretation of these limits: a heavy Higgs arising from an additional singlet scalar, which we present in Section \ref{sec:singlet}, and a heavy Higgs arising from an additional electroweak doublet scalar, which we present in Section \ref{sec:doublet}. In Section \ref{sec:twinhiggs} we explore a UV completion of the singlet-like Higgs model in the form of the fraternal Twin Higgs, and illustrate the prospects of LLP searches as a leading discovery channel in this setting. We conclude in Section \ref{sec:conclusions}, and reserve both the details of our reinterpretation procedure for LHC LLP searches and a summary of key properties of the heavy fraternal Twin Higgs boson for a trio of appendices.



\section{Direct searches for a scalar resonance}\label{sec:modelInd}

We begin by exploring the phenomenology of a second Higgs-like scalar with potentially exotic decays in a relatively model-independent fashion. We focus primarily on the case of a CP-even neutral scalar, which may be an isolated state or a component of a larger electroweak multiplet. In general, such a scalar resonance $\phi$ could have nonzero branching ratios directly into Standard Model states as well as into new states in a ``dark'' sector. The latter decays can lead to invisible and/or displaced signatures of a heavy scalar at the LHC. While the dark sector might contain a great diversity of new states, for simplicity we consider decays into a single dark sector state $X$, which we assume to be lighter than the scalar resonance: $m_X\ll m_\phi$.  

The two most relevant parameters to describe the phenomenology of this simplified model are the mass of the scalar $m_\phi$ and its signal strength in a given decay channel $\sigma_{\phi}\cdot\text{BR}$. While there are a variety of possible production modes for such a resonance, we will (for the most part) assume that gluon fusion is the dominant production mode at the LHC. Depending on the details of the model, the relative contribution of associated production modes will typically be less than or equal to that of a Standard Model Higgs of the same mass (see \cite{MelladoGarcia:2150771} for a summary of the different cross sections at different masses and center of mass energies). Among all the possible visible and displaced channels, we focus our attention on visible decays into di-boson and di-higgs final states -- which provide the leading constraints in heavy Higgs scenarios in the absence of parametrically enhanced couplings to specific flavors of quarks and leptons -- and on decays into $X$ bosons with lifetimes ranging from nearly prompt to detector-stable. While the $X$ boson may possess a variety of decay products, in what follows we will consider decays predominantly into pairs of $b$ quarks, as is to be expected from LLPs that inherit Standard Model couplings by mixing with the Higgs.
Of course, in such a scenario the LLP decays would begin to favor electroweak gauge bosons and eventually top quarks for $m_X \gtrsim 2 m_W$. Hadronic final states in these channels would lead to bounds qualitatively similar to those obtained from LLP decays into $b$ quarks.

Another two parameters of central importance are the mass $m_X$ of the dark sector state and its total decay width $\Gamma_X$ into SM states (henceforth assuming, for simplicity, that $X$ decays exclusively into the SM). These parameters determine the average decay length in the lab frame, which is of the form
\begin{equation}
c\tau_{\text{LAB}}\simeq c\tau_X\cdot\frac{m_\phi}{2 m_X}
\end{equation}
where $\tau_X = 1/\Gamma_X$ is the proper lifetime of $X$. The mass ratio between the mother particle $\phi$ and the daughter particle $X$ has also non-trivial consequences for the shape of the final event. This is mainly related to the fact that the angular opening of the decay products of the daughter particle $X$ is proportional to its boost in the lab frame, 
\begin{equation} 
\Delta R_{jj}\simeq \frac{4m_X}{m_\phi} \ .\label{eq:angularsep}
\end{equation}
This impacts the sensitivity of LLP searches given that many of them require the decay products of the LLP to be resolved.
All in all, the most relevant parameters for a model-independent characterization of a heavy Higgs at the lifetime frontier are
\begin{equation}
m_\phi\quad,\quad \sigma_{\phi}\cdot\text{BR}\quad, \quad m_X\quad\ ,\quad c\tau_X=1/\Gamma_X\quad\ .
\end{equation}
In what follows we review the status of 13 TeV searches on this parameter space, and present projections for HL-LHC. For completeness we present our results on displaced searches at 8 TeV in Appendix~\ref{app:8TeVsearches}. 

\subsection{Status at 13 TeV}\label{sec:13TeVsummary}

\begin{figure}[t]
\includegraphics[width=.5\textwidth]{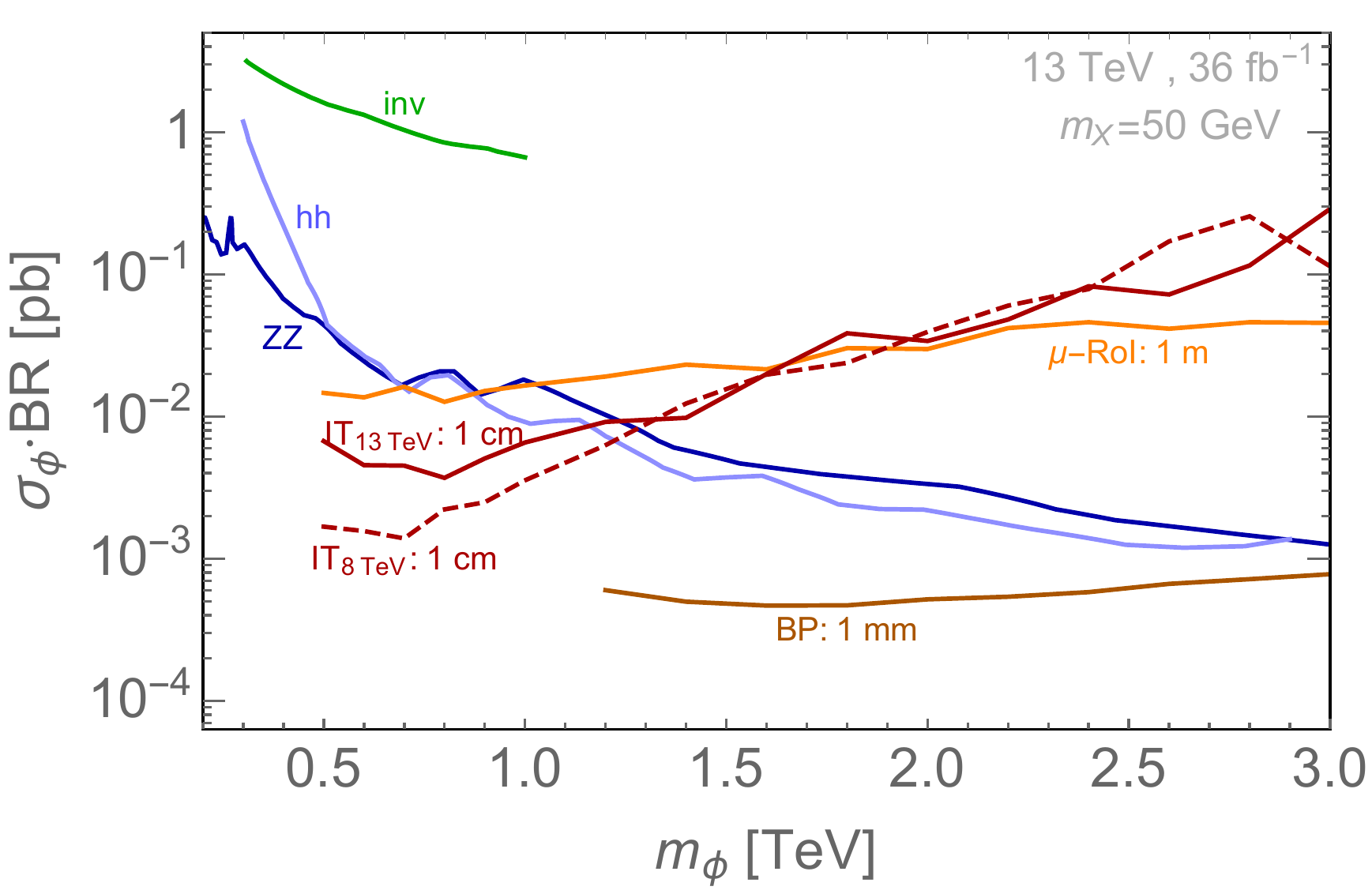}\hfill
\includegraphics[width=.5\textwidth]{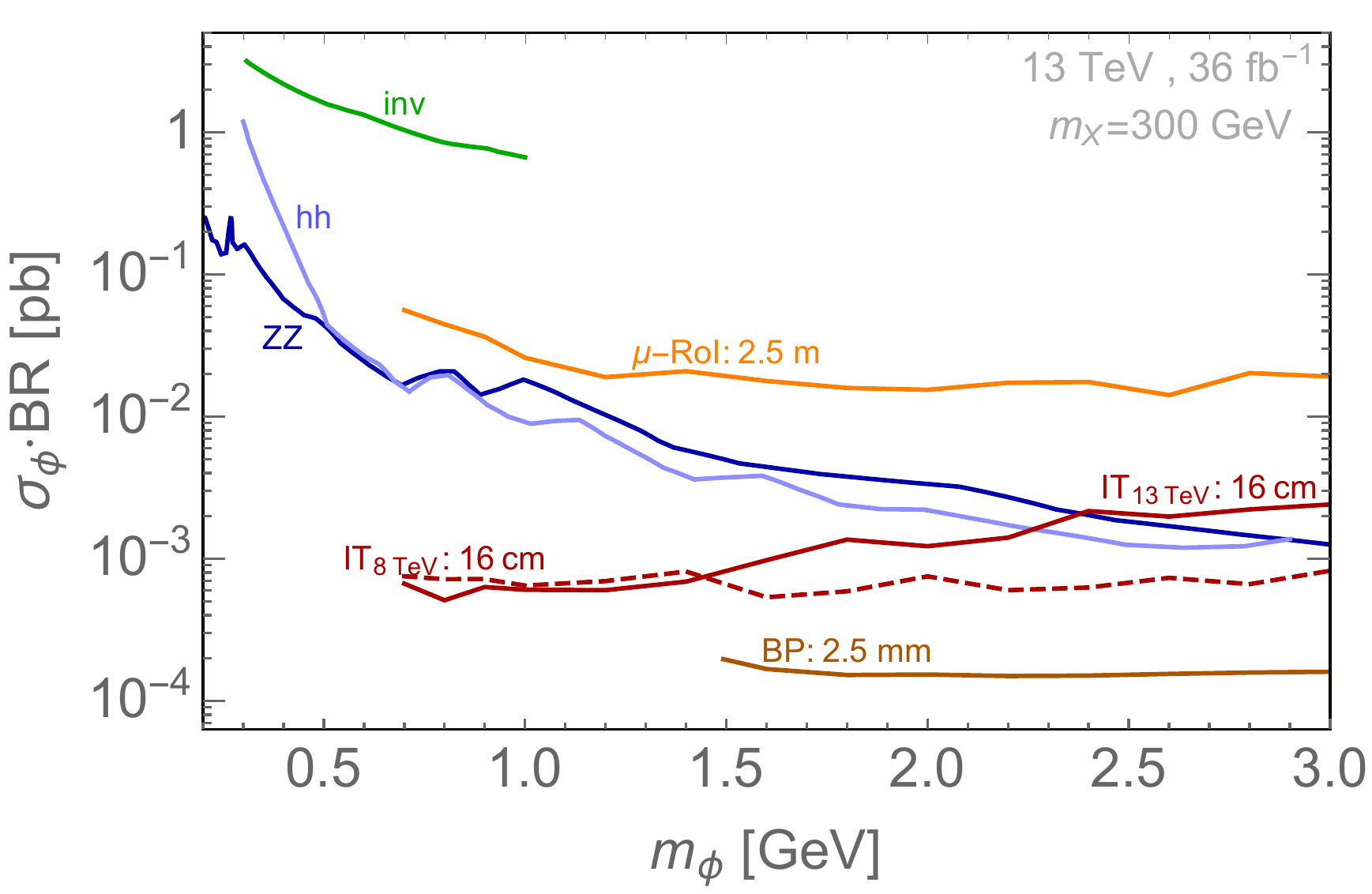}
\centering
\caption{Summary of the cross section times branching ratio probed by 13 TeV searches with $L_{\text{now}}=36\text{ fb}^{-1}$. {\bf Dark blue} for $\phi\to ZZ$ , {\bf dark green} for $\phi\to$ inv., {\bf light blue} for $\phi\to hh$. For $\phi\to XX$ we consider $~X\to jj$ with different values of the $X$ lifetimes and {\bf left} and {\bf right} panels correspond to $m_X=50\text{ GeV}$ and $m_X=300\text{ GeV}$, respectively. 
In the plot we select three representative values for these lifetime regimes which maximize the reach of a given search. 
{\bf Orange} is the ATLAS $\mu$-RoI  exclusion with $c\tau_X\sim \text{m}$ \cite{Aaboud:2018aqj}, {\bf dark red} is the CMS IT exclusion with $c\tau_X\sim \text{cm}$ \cite{Sirunyan:2018vlw}, {\bf dark orange} is the CMS beam pipe search with $c\tau_X\sim \text{mm}$ \cite{Sirunyan:2018pwn}. For comparison, we also show in {\bf dark red-dashed} the CMS IT exclusion at 8 TeV \cite{CMS:2014wda} with the same $c\tau_X$ projected with 13 TeV luminosity. The different data selection requirements of the 8 TeV IT search makes it more sensitive to scenarios with light $m_X$. The CMS BP search is effective for sufficiently high $m_\phi$ to pass the stringent $H_T$ requirement, which then necessitates larger $m_X$ in order to ensure the decay products remain resolved.}
\label{fig:xsecSummaryPlot}
\end{figure}

\begin{figure}[t]
\includegraphics[width=.33\textwidth]{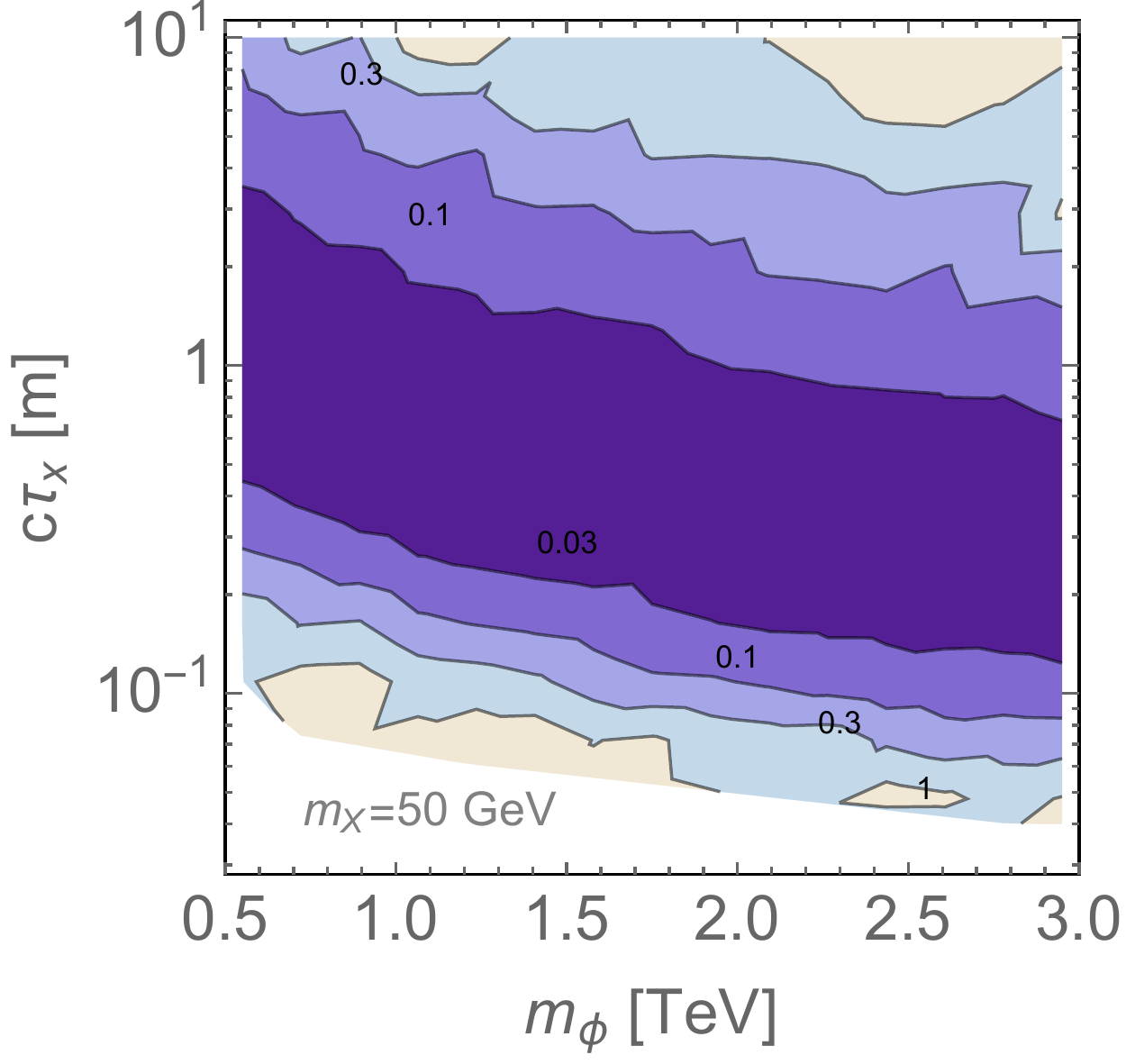}\hfill
\includegraphics[width=.33\textwidth]{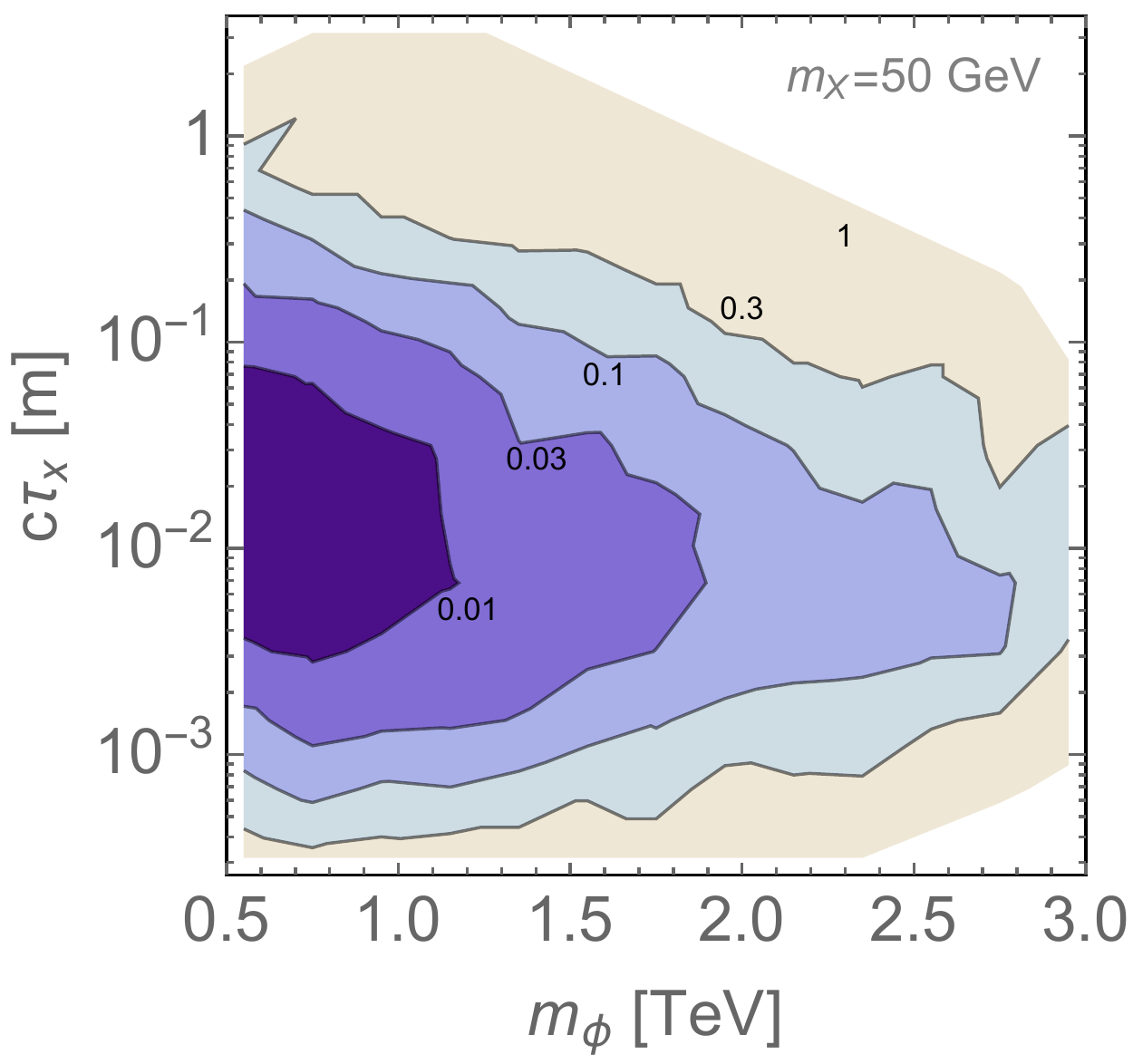}\hfill
\includegraphics[width=.33\textwidth]{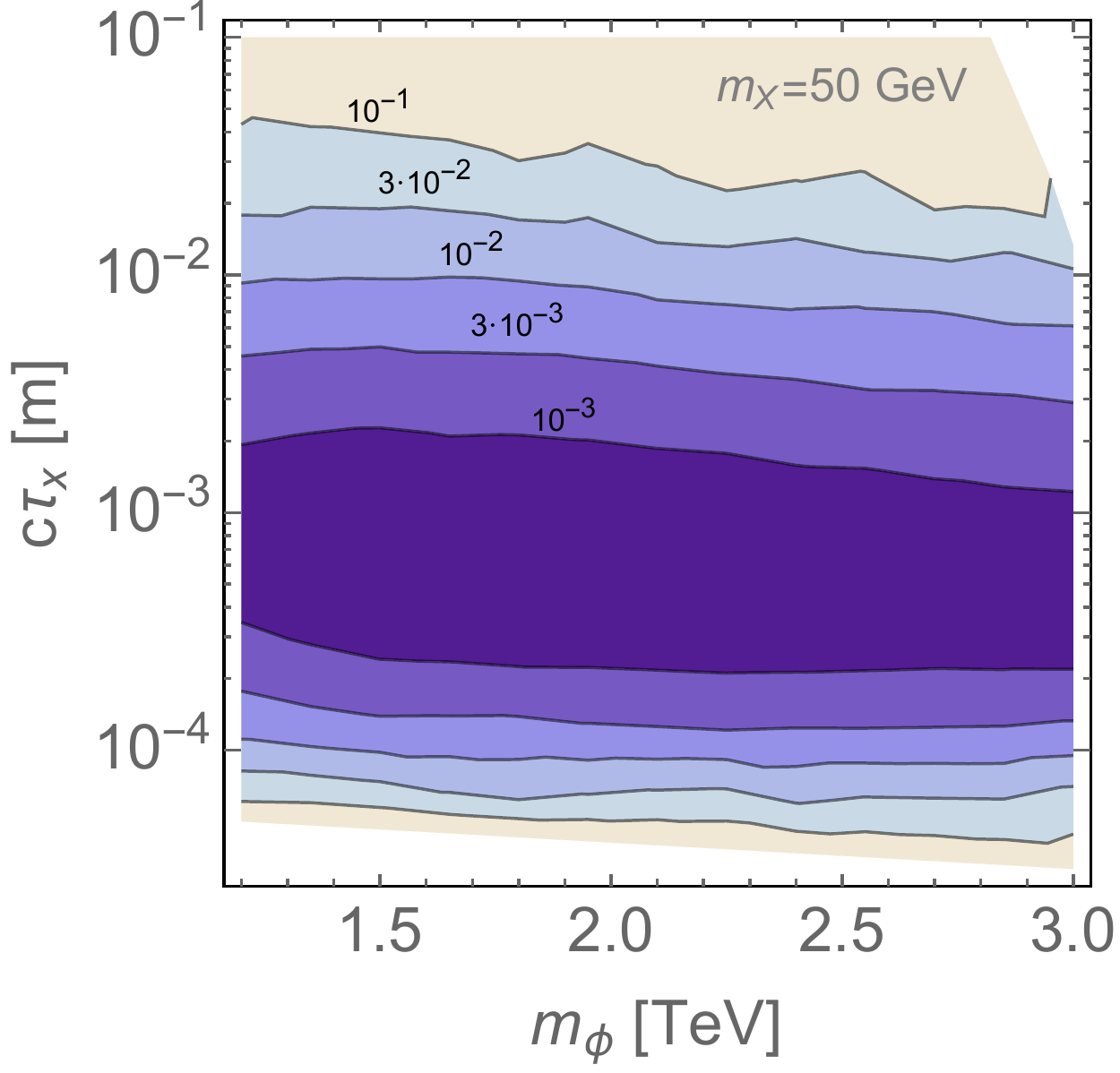}\\
\includegraphics[width=.33\textwidth]{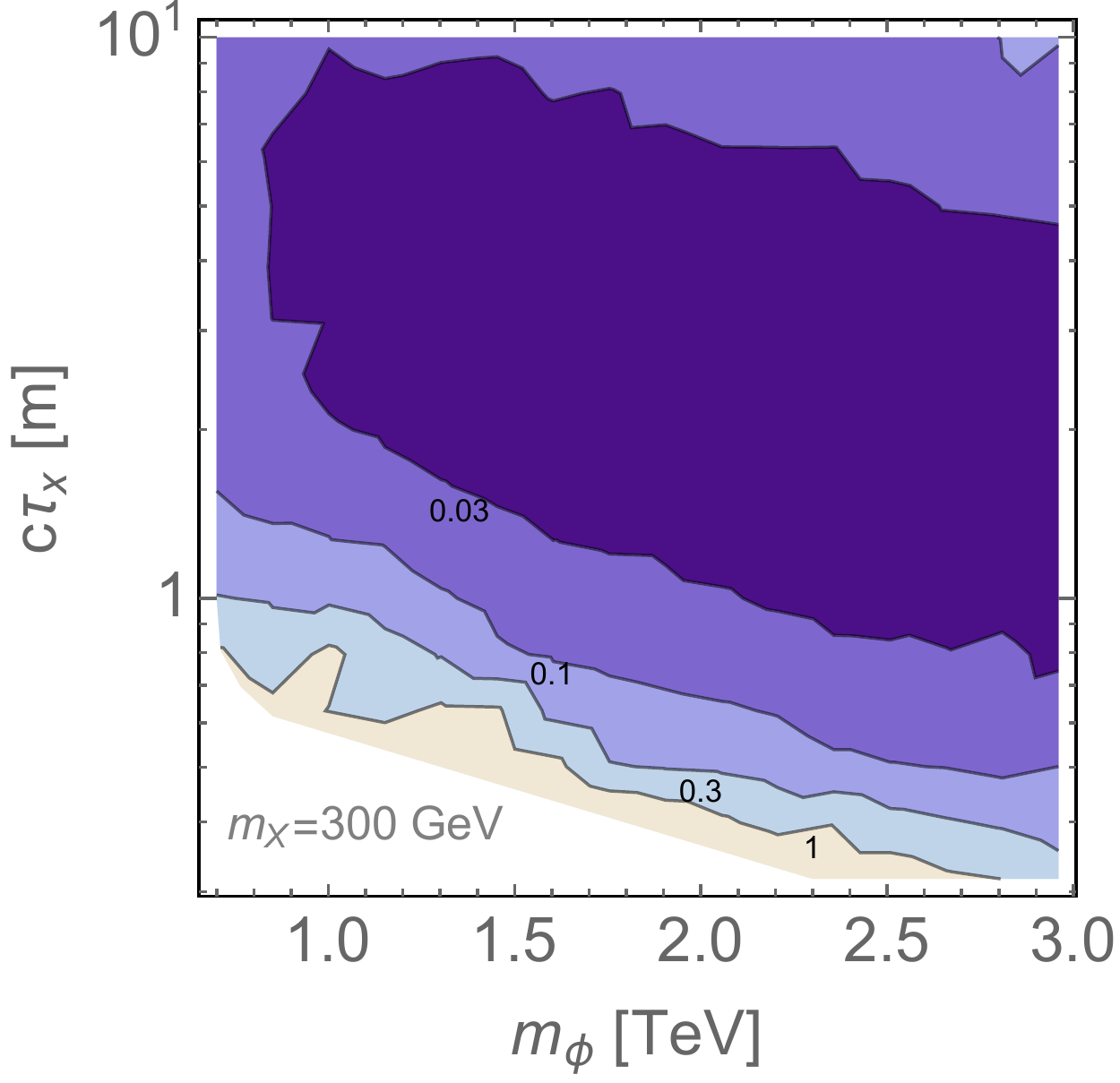}\hfill
\includegraphics[width=.33\textwidth]{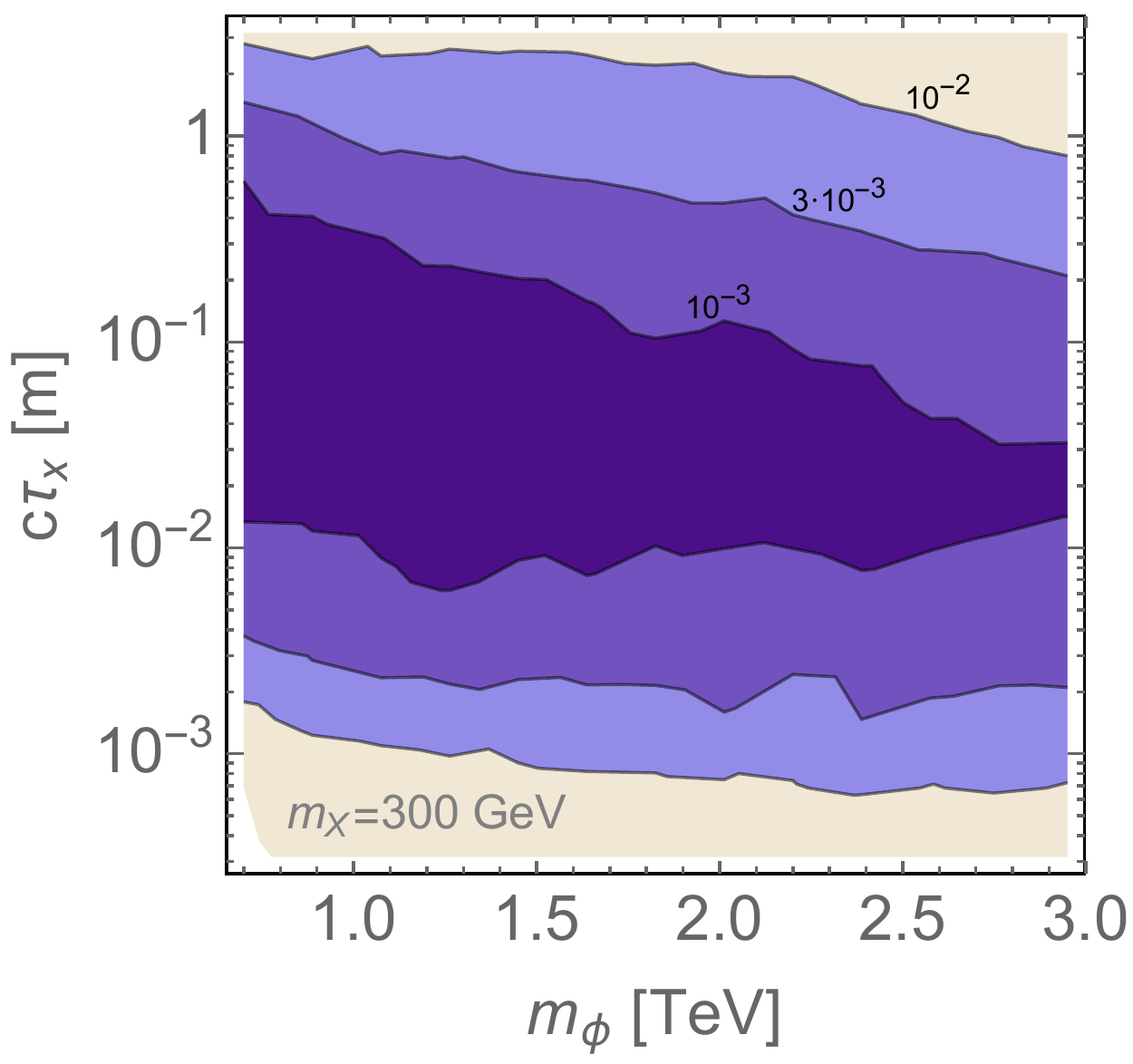}\hfill
\includegraphics[width=.33\textwidth]{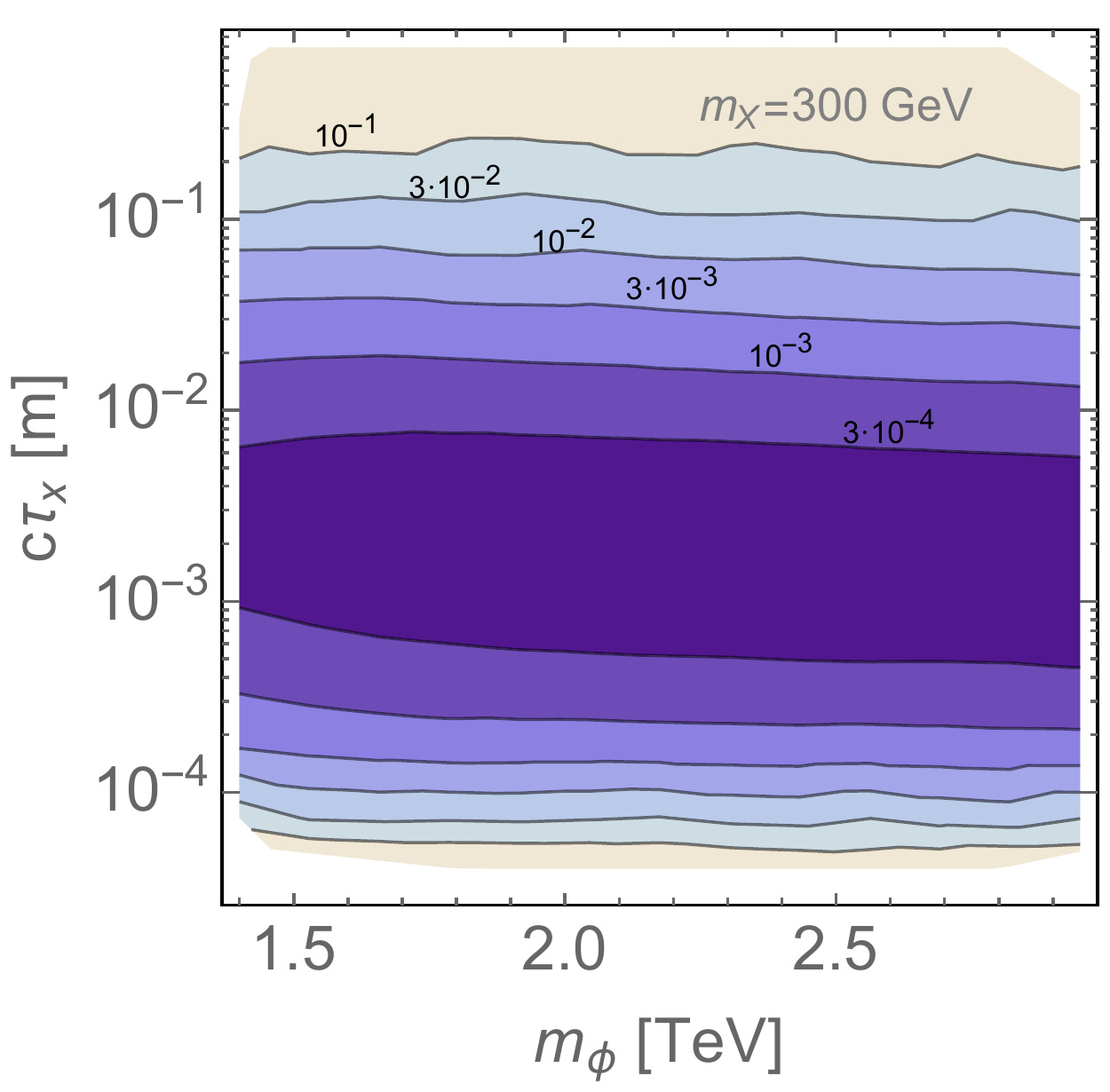}

\centering
\caption{Contours of the excluded signal strength $\sigma_{\phi}\cdot \text{BR}$ in pb in the plane ($m_\phi$, $c\tau_X$) for the three LLP searches discussed in the text. The upper/lower rows refer to $m_X=50\text{ GeV}$/$m_X=300\text{ GeV}$ respectively. {\bf Left:} ATLAS $\mu$-RoI search at 13 TeV \cite{Aaboud:2018aqj} {\bf Center:} CMS IT search at 13 TeV~\cite{CMS:2014wda}. {\bf Right:} CMS beam pipe search at 13 TeV~\cite{Sirunyan:2018pwn}. }
\label{fig:13TeVLLP}
\end{figure}

In Fig.~\ref{fig:xsecSummaryPlot} we present a set of relevant limits on the cross section times branching ratio, $\sigma\cdot \text{BR}$, coming from existing ATLAS and CMS searches in diverse final states at $\sqrt{s} =$ 13 TeV with $36\text{ fb}^{-1}$. For direct decays we show limits coming from the CMS $ZZ$ search based on a combination of the $4l$, $2l+2\nu$ and $2l+2q$ channels \cite{Sirunyan:2018qlb} and the ATLAS search in $hh$ further decaying into $4b$ \cite{Aaboud:2018knk}. Similar bounds can be obtained from analogous ATLAS/CMS searches \cite{Sirunyan:2017isc,ATLAS:2017spa}. We also include the invisible search at CMS \cite{CMS-PAS-HIG-17-023}, which assumes that the ratio of associated production to gluon fusion is comparable to a Standard Model Higgs of the same mass. As we will see, this assumption may be readily violated in simple models.

As for displaced searches at 13 TeV, we show representative limits from the following three reinterpretations: i) ATLAS 13 TeV search using the muon Region of Interest trigger ($\mu$-RoI) \cite{Aaboud:2018aqj} ii) CMS search using the inner tracker (IT) trigger at 13 TeV~\cite{CMS:2014wda} and iii) CMS search  using the beam pipe (BP) trigger at 13 TeV~\cite{Sirunyan:2018pwn}. We show that the combination of these three searches provide good coverage for different values of $X$ lifetimes ranging from meters ($\mu$-RoI trigger), to centimeters  (IT trigger) and millimeters (BP trigger). While in Fig.~\ref{fig:xsecSummaryPlot} we set the lifetime to three different representative values, in Fig.~\ref{fig:13TeVLLP} we show the sensitivity of the three displaced searches as a function of the heavy resonance mass $m_\phi$ and the daughter lifetime $c\tau_X$ for two different choices of daughter mass $m_X=50,\,300\text{ GeV}$ (upper and lower panels, respectively). The details of these three searches and our corresponding reinterpretations are as follows:
 
\begin{itemize}
\item The ATLAS 13 TeV search using the $\mu$-RoI trigger \cite{Aaboud:2018aqj}, is an update of the previous 8 TeV analysis~\cite{Aad:2015uaa} searching for displaced hadronic jets appearing in the muon chamber. The muon Region of Interest (RoI) trigger around which this search is designed is tailored to tag displaced decays with decay length $0.5\text{ m}\lesssim c\tau\lesssim 20\text{ m}$ \cite{Aad:2013txa} where the muon reconstruction algorithm is fully efficient (see Ref.~\cite{Aad:2013ela}). Our reinterpretation procedure is qualitatively similar to the 8 TeV reinterpretation which we present in Appendix~\ref{app:8TeVsearches}, but uses updated information about the trigger and vertex reconstruction efficiency appropriate to the 13 TeV search. The 95\% C.L. exclusion limit is given by 
\begin{equation}
\sigma_\phi^{13\text{ TeV}}\cdot\text{ BR}=  0.083\text{ fb}\cdot \frac{L}{36.1\text{ fb}^{-1}}\cdot \frac{1}{\epsilon(m_\phi,m_X,c\tau_X)}\ ,
\end{equation}
where $\epsilon(m_\phi,m_X,c\tau_X)$ accounts for the detector acceptance and efficiency for the signal and, for convenience, we include a possible luminosity scaling by an integrated luminosity $L$ relative to the actual integrated luminosity used in the search. Our simulation procedure for obtaining $\epsilon(m_\phi,m_X,c\tau_X)$ is summarized in Appendix \ref{app:ATLASmuon}. The exclusion power of this search is comparable to the ones from visible decays for optimal displacements ($ c\tau\sim 1$ m) and sub-TeV masses (see Fig. \ref{fig:xsecSummaryPlot}), and it deteriorates for longer and shorter displacements. As we can see from the left panel of Fig.~\ref{fig:13TeVLLP}, in a large region of $c\tau$ the reach of this search is independent of the mass of the singlet, $\phi$, once $c\tau$ and $m_X$ are fixed. The residual dependence at the boundaries can be easily explained by considering the extra boost factor of the $X$ in the lab-frame coming from the decay of an heavier singlet. Heavier $m_\phi$ can increase the sensitivity for $c\tau_X\lesssim 1\text{m}$ and decrease it for  $c\tau_X\gtrsim 1\text{m}$. By comparing the sensitivity at 13 TeV with the 8 TeV one (see Appendix~\ref{app:8TeVsearches}) we can see how a LLP search at the LHC can be zero-background: the 13 TeV muon RoI analysis remains background-free with trigger performance comparable to the 8 TeV analysis. 

\item The CMS 13 TeV search using the Inner Tracker trigger \cite{Sirunyan:2018vlw} updates the previous 8 TeV analysis~\cite{CMS:2014wda} for displaced dijet pairs appearing in the inner tracker. Both at 8 TeV and at 13 TeV this search is nearly background free. Our simulation procedure for obtaining $\epsilon(m_\phi,m_X,c\tau_X)$ is summarized in Appendix~\ref{app:CMS8TeV}-\ref{app:CMS13TeV} together with a validation of our recasting. The 95\% C.L. exclusion limit is given by
\begin{equation}
    \sigma_\phi^{13\text{ TeV}}\cdot\text{ BR}= 0.1156\text{ fb}\cdot \frac{L}{35.9\text{ fb}^{-1}}\cdot \frac{1}{\epsilon(m_\phi,m_X,c\tau_X)}
\end{equation}
From Fig.~\ref{fig:xsecSummaryPlot} we see that the exclusion power strongly depends on the hierarchy between the mother resonance mass, $m_\phi$, and the daughter resonance mass, $m_X$. For $m_X=50\text{ GeV}$ the exclusion powers depends strongly on the mass of the mother resonance $m_\phi$, which controls the boost of $X$ and the angular separation between the jet pairs. It is interesting to notice that the selection criteria of the 8 TeV analysis~\cite{CMS:2014wda} allows for a larger efficiency at large boost factor. For this reason, in Fig.~\ref{fig:xsecSummaryPlot} we plot the expected reach of the 8 TeV search rescaled linearly with the luminosity of the other 13 TeV searches: $L_{8\text{TeV}}/L_{13\text{TeV}}\simeq0.91$. More details about this comparison will be given in Appendix~\ref{app:8TeVsearches}. Comparing the upper and lower middle panels of Fig. \ref{fig:13TeVLLP}, we can see that the large dependence on the mother mass is reduced for $m_X=300\text{ GeV}$. Also the difference between the efficiency of the 8 TeV and the 13 TeV search is reduced. A slight degradation due to the $X$ boost can only be seen at high $m_\phi$.

\item The CMS 13 TeV search looking for two displaced vertices in the beampipe (BP)~\cite{Sirunyan:2018pwn} provides a straightforward generator-level recipe for obtaining limits on new interpretations that is faithful for signal benchmarks with high acceptance, which we discuss in Appendix \ref{app:CMSbp}. The 95\% C.L. exclusion limit is given by 
\begin{equation}
\sigma_\phi^{13\text{ TeV}}\cdot\text{ BR}= 0.078\text{ fb}\cdot \frac{L}{38.5\text{ fb}^{-1}}\cdot \frac{1}{\epsilon(m_\phi,m_X,c\tau_X)}
\end{equation}
This analysis is only effective for $m_\phi \gtrsim 1$ TeV due to the substantial $H_T$ requirement, and is mostly sensitive to heavy $m_X$. This is due to the requirement that both displaced vertices be associated with a resolved jet pair, so that the search efficiency drops rapidly once the average separation of the decay products is boosted to within the size of the jet radius. This behavior appears very clearly by comparing the upper and lower right panels of Fig.~\ref{fig:13TeVLLP}. When the  $H_T$ requirement is satisfied, this search covers the region of very short displacement with efficiencies maximized for $c\tau_X\simeq \text{ mm}$.

\end{itemize}

\subsection{Projections at HL-LHC} 

\begin{figure}[t]
\includegraphics[width=.5\textwidth]{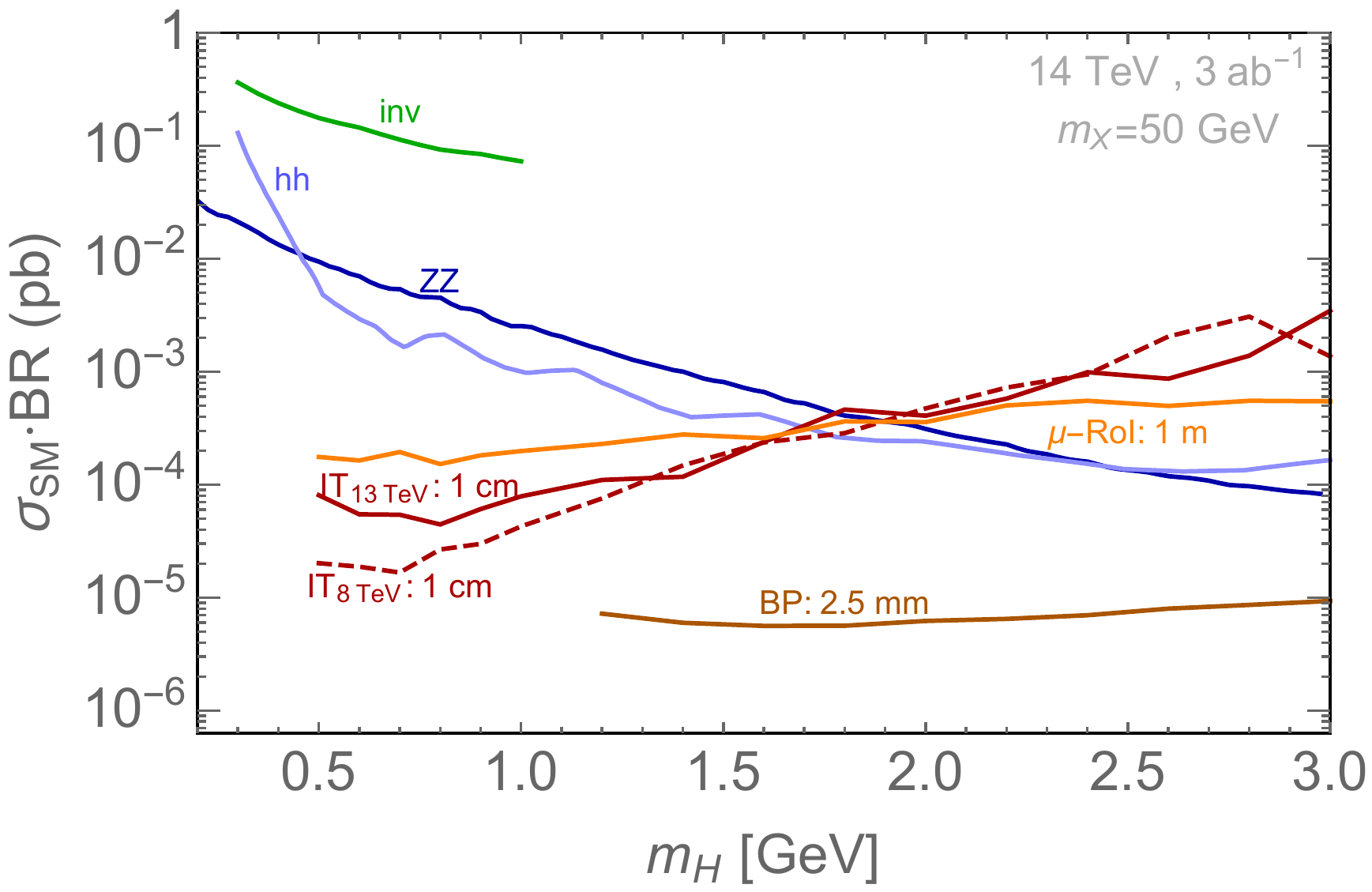}\hfill
\includegraphics[width=.5\textwidth]{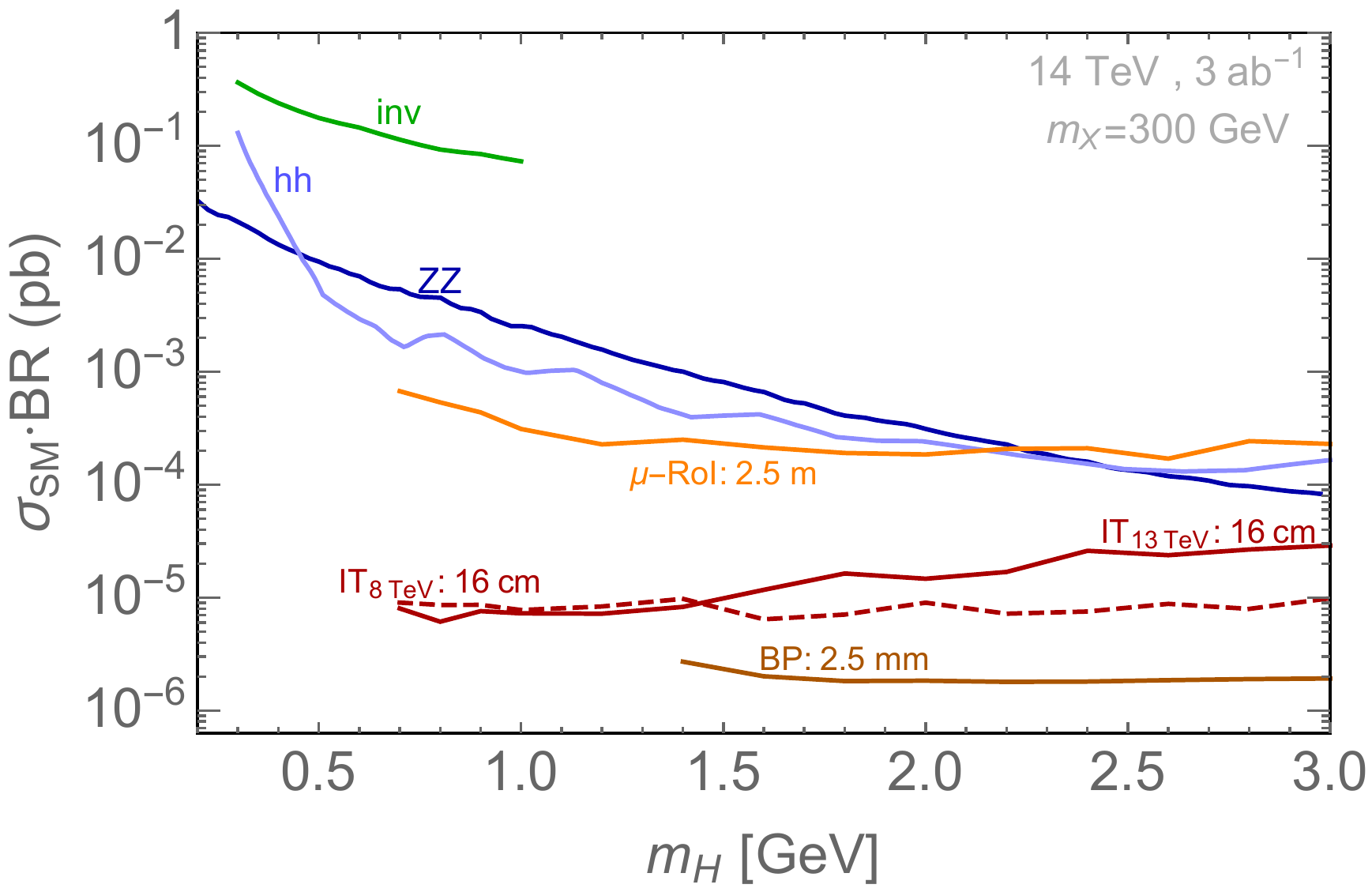}
\centering
\caption{Projected reach in cross sections  times branching ratio at HL-LHC. {\bf Dark blue} for $\phi\to ZZ$, {\bf dark green} for $\phi\to$ inv., {\bf light blue} for $\phi\to hh$. These channels are rescaled by $\sqrt{L_{\text{HL-LHC}}/L_{\text{13 TeV}}}$. For $\phi\to XX$ we consider $~X\to jj$ with different values of the $X$ lifetimes and {\bf left} and {\bf right} panels correspond to $m_X=50\text{ GeV}$ and $m_X=300\text{ GeV}$. For these channels, we rescale the results in Fig.~\ref{fig:xsecSummaryPlot} with $L_{\text{HL-LHC}}/L_{\text{13 TeV}}$. {\bf Orange} is the ATLAS $\mu$-RoI  exclusion with $c\tau_X\sim \text{m}$ \cite{Aaboud:2018aqj}, {\bf dark red} is the CMS IT exclusions at 13 TeV $c\tau_X\sim \text{cm}$ \cite{Sirunyan:2018vlw}, {\bf dark red-dashed} is the CMS IT exclusion at 8 TeV \cite{CMS:2014wda} with the same $c\tau_X$ projected with 13 TeV luminosity, and {\bf dark orange} is the CMS beam pipe search at 13 TeV with $c\tau_X\sim \text{mm}$ \cite{Sirunyan:2018pwn}. 
\label{fig:xsecHLLHC}}
\end{figure}

In Fig.~\ref{fig:xsecHLLHC}, we show the extrapolated reach for the high-luminosity phase of the LHC with $L_{\text{HL-LHC}}=3\text{ ab}^{-1}$. For the visible searches we use the rescaling procedure already used in \cite{weilersalam,Thamm:2015zwa,Buttazzo:2015bka}. Since the parton luminosities controlling the background do not change drastically between 13 TeV and 14 TeV, the rescaling is essentially controlled by the squared root of the ratio of luminosities $\sqrt{L_{\text{now}}/L_{\text{HL-LHC}}}\simeq 0.1$. The same rescaling is applied to the invisible searches at 13 TeV.

As far as displaced searches are concerned, we rescale the bounds linearly with the luminosity, assuming their background to remain constant at higher luminosity. This aggressive extrapolation is to some extent already supported by the scaling of the background of the muon RoI search between the 8 TeV dataset \cite{Aad:2015uaa} and the 13 TeV dataset \cite{Aaboud:2018aqj}. Indeed with very few changes in the actual search from 8 TeV to 13 TeV, the number of background events at 13 TeV is consistent with zero while the luminosity increases by a factor of 2. Of course a variety of new challenges to the background characterization of LLP searches are expected to arise at high luminosity, making this extrapolation optimistic. That said, additional hardware and trigger improvements (such as track triggers  \cite{Gershtein:2017tsv} and precision timing \cite{Liu:2018wte}) are likely to keep pace.

Let us finally comment on the main result of this model-independent parameterization. As one can see from Fig.~\ref{fig:xsecHLLHC}, displaced searches at the HL-LHC have an unprecedented discovery potential for new heavy SM-like resonances. This is due to the very low backgrounds of these searches compared to the usual search strategies based on visible decays. In explicit beyond-the-Standard Model scenarios, the reach of these searches can often compensate for the reduced signal rates of these exotic decays compared to the visible channels.  As far as the lifetime coverage is concerned, it is clear that different search strategies based on different detector components can be sensitive to a wide range of displacements ranging from the sub-centimeter level (as in the CMS beam-pipe analysis \cite{Sirunyan:2018pwn}), to centimeters (as in the CMS inner tracker analysis \cite{CMS:2014wda}), to meters (as in the ATLAS muon chamber analysis \cite{Aad:2015uaa,Aaboud:2018aqj}). 

An interesting  observation is the degradation of most of the displaced search strategies for higher masses of the mother resonance, for relatively light daughter particles. This is mainly due to the collimation of the daughter decay products (see Eq.~\eqref{eq:angularsep}), and is generic, as far as we expect the dark sector states to be lighter than the second Higgs. A notable exception is the  ATLAS muon chamber analysis \cite{Aad:2015uaa,Aaboud:2018aqj} which indeed is likely to give the best reach for heavier resonances. The natural next step to extend the coverage of other searches to heavier resonance masses would be to use jet-substructure techniques to resolve the collimated tracks coming from the boosted $X$ decays.

\section{Displaced decays of a singlet Higgs}\label{sec:singlet}

While the model-independent bounds presented in the previous section provide a useful guide to the relative strength of prompt and displaced searches for a second Higgs, a true comparison is only possible in the context of models that relate the production and decays of the Higgs. Here we present the first of two such models, in which the second Higgs arises from a real CP-even scalar, $S$, that couples to the Standard Model via the Higgs portal.

We introduce the effective lagrangian of a CP-even scalar up to dimension four:
\begin{equation}
\mathcal{L}_{\text{visible}}=\frac{1}{2}(\partial_\mu S)-\frac{1}{2} m_S^2 S^2-a_{HS} S\vert H\vert^2-\frac{\lambda_{HS}}{2} S^2\vert H\vert^2-\frac{a_S}{3} S^3-\frac{\lambda_S}{4} S^4\ .\label{eq:everybody}
\end{equation}
After electroweak symmetry breaking, the singlet mixes with the uneaten CP-even component of the Higgs doublet. In the limit that the mixing is small and the singlet mass parameter is larger than the Higgs doublet mass parameter, we can express the mixing angle, $\gamma$, as
\begin{equation}
\gamma\simeq\frac{v(a_{HS}+\lambda_{HS} f)}{m_\phi^2}\quad\ ,\qquad H=\begin{pmatrix}\pi^+\\ \frac{v+h}{\sqrt{2}}\end{pmatrix}\quad\ ,\qquad S=f+\phi\ ,
\end{equation}   
where $m_\phi$ is the physical mass of the singlet, $v=246\text{ GeV}$ is the electroweak vacuum expectation value (VEV), and $f$ is the VEV of the singlet $S$. In what follows we will often abuse terminology and refer to the mass eigenstates as the ``singlet Higgs" and ``SM-like Higgs", though of course they are ultimately (modest) admixtures.

The above formula shows how the mixing between the singlet and the SM Higgs is controlled by the spontaneous and/or explicit breaking of a discrete $\mathbb{Z}_2$ symmetry under which the singlet is odd ($S\to -S$) and the SM Higgs even $(H\to H)$. Parametrically, we can distinguish three scenarios: 
\begin{enumerate}
\item The singlet takes a VEV at the minimum of a $\mathbb{Z}_2$-invariant potential. Then $m_\phi^2\simeq 3\lambda_S f^2$, the $\mathbb{Z}_2$-breaking is spontaneous, and the mixing with the SM Higgs is approximated by 
\begin{equation}
\gamma\simeq \frac{\lambda_{HS}}{\lambda_S}\cdot\frac{v}{f}\ .
\end{equation}
\item The primary source of  $\mathbb{Z}_2$-breaking is the explicit breaking due to the singlet trilinear coupling with the SM Higgs. In this case the mixing goes as 
 \begin{equation}
 \gamma\simeq\frac{ a_{HS} v}{m_\phi^2}\ ,
 \end{equation} 
 and can be made arbitrarily small. We refer to \cite{Barbieri:2006bg} for a discussion of explicit models where this scaling is realized. 
 
 \item If the $\mathbb{Z}_2$-symmetry is exact, the singlet can only be pair produced at colliders and the Higgs couplings are modified only at loop level. This is the so-called ``nightmare scenario'' which presents interesting phenomenological challenges and could provide a minimal scenario for EW baryogenesis \cite{Curtin:2014jma}. 
 \end{enumerate}
 
 In what follows we will focus mainly on the first scenario. This is explicitly realized in Twin Higgs scenarios where $\lambda_S\simeq \lambda_{HS}$ and $\gamma\sim \frac{v}{f}$ (see Refs~\cite{Chacko:2005pe, Barbieri:2005ri, Buttazzo:2015bka, Chacko:2017xpd, Ahmed:2017psb}). From Eq.~\eqref{eq:everybody} the phenomenology of the singlet and the SM-like Higgs is completely controlled by the mixing angle $\gamma$ and can be summarized as follows:
\begin{align}
&\frac{g_{hVV,f\bar f}}{g_{hVV,f\bar f}^{SM}}=\cos^2\gamma\,, \label{eq:SMHiggs}\\
&\sigma_\phi=\sin^2\gamma\cdot \sigma_{h}(m_\phi)\,,\label{eq:Sproduction}\\
&\text{BR}_{\phi\to f\bar{f},VV}=\text{BR}_{h\to f\bar{f},VV}(1-\text{BR}_{\phi\to hh \label{eq:SBR}
})\,.
\end{align}
Here $g_{hVV}/g_{hVV}^{SM}$ and $g_{hf\bar f}/g_{hf\bar f}^{SM}$ refer to the couplings of the SM-like Higgs to Standard Model vectors and fermions, respectively, normalized to the Standard Model prediction. The cross section $\sigma_h(m_\phi)$ is that of a Standard Model Higgs of mass $m_\phi$. From Eq.~\eqref{eq:SMHiggs} we see that the couplings of the SM-like Higgs to SM states are reduced by $\cos^2\gamma$, leading to a reduced production cross section in every channel but unchanged branching ratios.  Eq.~\eqref{eq:Sproduction} shows that the production cross section of the heavy singlet is the one of the corresponding Higgs boson at mass $m_\phi$ rescaled by $\sin^2\gamma$.  Finally, the branching ratios of the singlet into SM gauge bosons in Eq.~\eqref{eq:SBR} are rescaled by a common factor depending on the branching ratio into $hh$. The latter is in principle model dependent, however, for $m_\phi\gg m_W$ the potential has an approximate $SO(4)$ symmetry which implies 
\begin{equation}
\text{BR}_{\phi\to hh }\simeq \text{BR}_{\phi\to ZZ}\simeq \text{BR}_{\phi\to WW} /2\ .
\end{equation}

\begin{figure}[t]
\includegraphics[width=.5\textwidth]{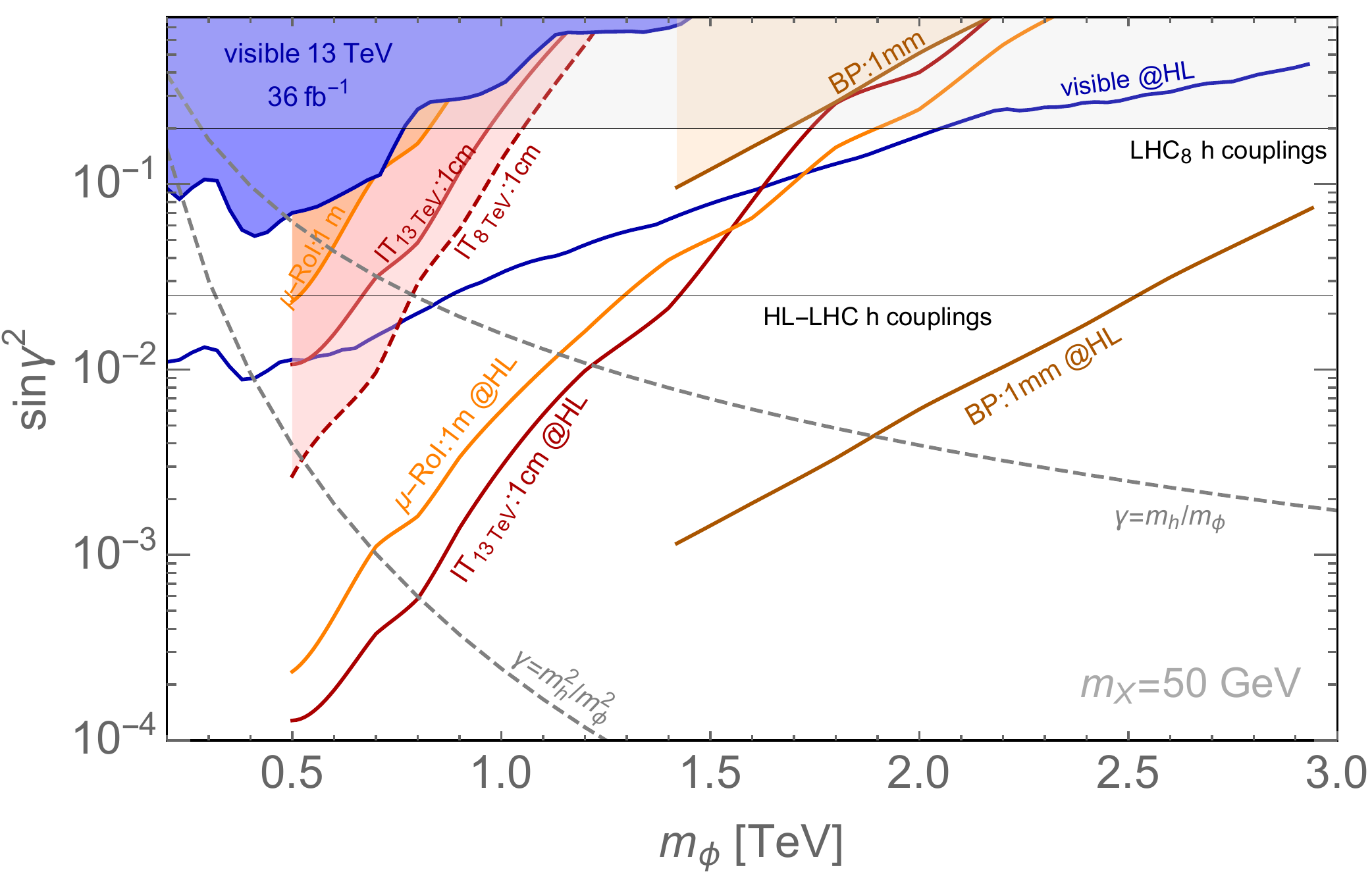}\hfill
\includegraphics[width=.5\textwidth]{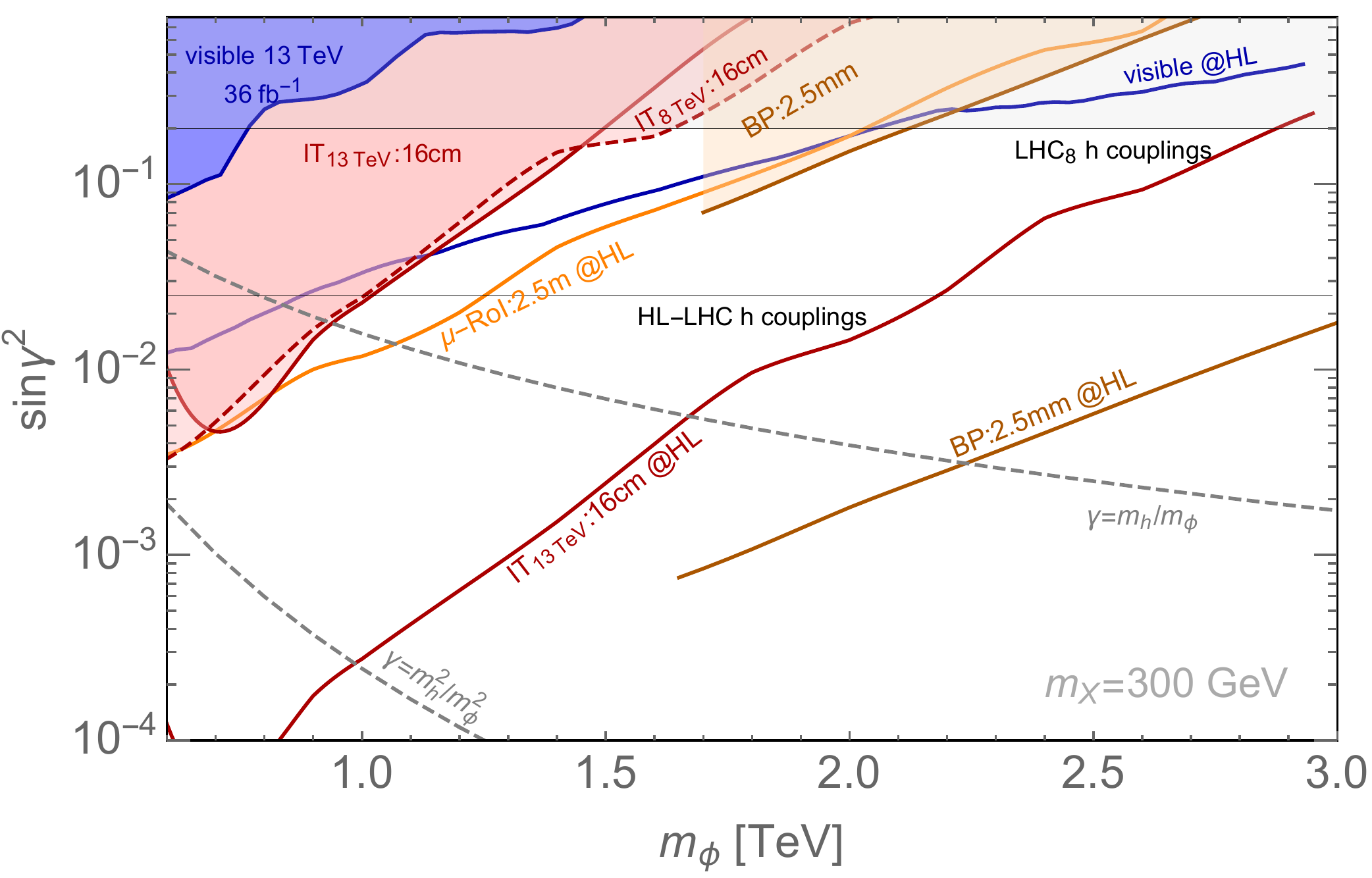}
\centering
\caption{Parameter space of the singlet Higgs as a function of $m_\phi$ and $\sin^2 \gamma$ overlaid with current and projected constraints from direct searches, as well as current and projected indirect limits from coupling measurements of the Standard Model-like Higgs at 125 GeV. We choose $m_X=50\text{ GeV}$ in the {\bf left} panel, and $m_X=300\text{ GeV}$ in the {\bf right} panel. The {\bf blue shaded} region is excluded by the combination of current searches for decays in the $ZZ$ and $hh$ final states at $\sqrt{s} = 13$ TeV, while the {\bf blue line} indicates the projected reach in the same channels for the HL-LHC. The {\bf grey shaded} region is excluded by the combination of ATLAS and CMS Higgs coupling measurements at $\sqrt{s} = 7$ \& 8 TeV, while the {\bf grey line} indicates projections for the corresponding reach at the HL-LHC. The two {\bf grey dashed lines} indicate natural parametric scalings of the mixing angle $\gamma$ with the mass of the singlet Higgs ($\gamma=m_h/m_\phi$ and $\gamma=m_h^2/m_\phi^2$). For the displaced decays we assume $\text{BR}_{\text{displaced}}\simeq \text{BR}_{\phi \rightarrow ZZ}=1/8$. The {\bf red shaded} region indicates the exclusion of the CMS inner tracker (IT) search at $\sqrt{s} = 13$ TeV assuming $c\tau_X=1\text{ cm}$ on the left panel and  $c\tau_X=16\text{ cm}$ on the right panel, the {\bf dashed red} line indicates the projection of the previous IT search at $\sqrt{s} = 8$ TeV for the same lifetimes. The {\bf red line} is the HL-LHC reach assuming zero background and same lifetimes. The {\bf orange shaded} region is excluded by the CMS beampipe (BP) search at $\sqrt{s} = 13$ TeV for $c\tau_X=1\text{ mm}$ on the left panel and $c\tau_X=2.5\text{ mm}$ on the right panel. The {\bf orange line} indicates the projected reaches for the HL-LHC assuming zero background and same lifetimes. The {\bf orange shaded} region on the left panel indicates the exclusion of the ATLAS muon RoI search at $\sqrt{s} = 13$ TeV for $c\tau_X=1\text{ m}$ while on the right panel the exclusion region for $c\tau_X=2.5\text{ m}$ is already covered by the visible search. The orange lines are the projected reaches for the HL-LHC assuming zero background and same lifetimes.  
\label{fig:singlet}
}
\end{figure}

The lagrangian in Eq.~\eqref{eq:everybody} has triggered intense phenomenological studies with the aim of comparing the reach of direct searches for a singlet decaying promptly to SM states, with the sensitivity from Higgs coupling measurements at the LHC and at future colliders \cite{Dawson:2013bba, Buttazzo:2015bka, Chacko:2017xpd, Ahmed:2017psb, Buttazzo:2018qqp}. 

We summarize in Fig.~\ref{fig:singlet} the relative strength of existing and future di-boson and di-higgs searches at the LHC, as well as constraints coming from the precision measurement of Higgs couplings (taking for definiteness the values in \cite{Cepeda:2019klc}).
In addition, we also show the possible reach of LHC displaced searches. In principle there are three qualitatively distinct branching ratios that determine the relative contribution of displaced searches: (1) the branching ratio into prompt or ``visible'' final states, $\text{BR}_{\text{visible}}$; (2) the branching ratio into long-lived or ``displaced'' final states, $\text{BR}_{\text{displaced}}$; (3) an additional branching ratio into detector-stable or ``invisible'' final states, $\text{BR}_{\text{invisible}}$. 
 In Fig.~\ref{fig:singlet} we assume $\text{BR}_{\text{displaced}} \simeq \text{BR}_{\phi \rightarrow ZZ}=1/8$ and $\text{BR}_{\text{invisible}} = 0$, for simplicity. 

The interplay of searches for visible decays, displaced decays, and Higgs coupling deviations highlights a notable feature of future LHC sensitivity to a singlet Higgs. In the absence of singlet Higgs decays into LLPs, the sensitivity of direct searches at the HL-LHC is unlikely to surpass limits from Higgs coupling measurements for $m_\phi \gtrsim 1.5$ TeV. However, for singlet Higgses decaying partly into LLPs, the potentially considerable reach of searches for displaced decays makes a direct search program competitive with Higgs coupling measurement to much higher values of $m_\phi$. The primary weakness of the displaced searches is at high $m_\phi$, low $m_X$, and large $c \tau$, where the muon RoI search loses sensitivity. Optimal coverage of this region could in principle be provided by MATHUSLA \cite{Curtin:2018mvb} or other proposed experiments.

Having constructed an explicit model for the heavy singlet Higgs, it is worth briefly exploring an explicit model for the LLP $X$ as well. In the minimal scenario of Eq.~\eqref{eq:everybody} one might naively conclude that having a displaced signature would always come at the price of a large suppression of the signal rate. Indeed since $\Gamma_{\text{visible}}\simeq \frac{\lambda_{HS}}{8\pi} m_\phi$, the only way of suppressing the decay width of the singlet itself is to suppress its mixing with the SM Higgs. 

The situation becomes drastically different when the singlet $S$ is itself a portal to a generic  dark sector. In this case the singlet $S$ can decay abundantly to a pair of dark states without suppressing the signal rate. Depending on the specific structure of the dark sector, these states can be approximately long lived and lead to displaced or invisible signatures for $S$. A simple example is to add an extra dark singlet scalar daughter $X$ to the Lagrangian in Eq.~\eqref{eq:everybody}:
\begin{equation}
\mathcal{L}_{\text{displaced}}=-\frac{a_{SX}}{2} S X^2-\frac{b_{SX}}{2} S^2 X-\frac{\lambda_{SX}}{4} S^2 X^2-\frac{\lambda_{SX}}{4} \vert H\vert^2 X^2-\frac{m_X^2}{2} X^2\ .
\end{equation} 
This type of setup arises naturally in Twin Higgs constructions \cite{Craig:2015pha} and it is further motivated by a class of Hidden Valley models where a rich and approximately stable hidden sector communicates with the Standard Model via a Higgs portal (see Refs~\cite{Strassler:2006ri, Han:2007ae}). If both $S$ and $X$ are odd under an approximate $\mathbb{Z}_2$-symmetry then $a_{SX}\simeq b_{SX}\simeq0$, and for $m_S>2 m_X$ the singlet $S$ will decay into pairs of dark sector states with a width $\Gamma_{\text{displaced}}=\frac{\lambda_{SX}^2 f^2}{8\pi m_S}$ which is now independent of the mixing of $S$ with the SM Higgs. The width of $X$ into SM states is instead proportional to the $\mathbb{Z}_2$-breaking parameters and can be arbitrarily suppressed.  

It is worth noticing that in this simple model, avoiding a fine-tuning of the mass of the scalar daughter $X$ gives an upper bound on $\Gamma_{\text{displaced}}$, which suppresses the rate of $\phi$ decays into dark states compared to the one into SM ones: $\Gamma_{\text{displaced}}/\Gamma_{\text{visible}}\lesssim \frac{\lambda_{SX}}{\lambda_{SH}}\cdot\frac{m_X^2} {m_\phi^2}$. This feature does not depend on the spin of the hidden sector state, as can be explicitly checked replacing the singlet $X$ with a vector or a fermion. This upper bound is easily circumvented if a cascade of decays of multiple states occurs in the hidden sector and/or if the UV contribution to the daughter mass is absent. In Sec.~\ref{sec:twinhiggs} we will see how the Twin Higgs gives an explicit realization of this simplified model where the singlet $X$ is identified with the lightest glueball.  The mass of the glueball is naturally lighter because of dimensional transmutation in the dark sector, and the decay of the singlet is unsuppressed because of the rich structure of the hidden sector where heavier states decay down to the lightest dark state.

\section{Displaced decays of a doublet Higgs}\label{sec:doublet}

Another simple scenario that gives rise to the decays of a second Higgs into LLPs arises when the Standard Model Higgs sector is extended by the addition of a new electroweak doublet scalar. We introduce the potential of a new Higgs doublet, $H_1$, with hypercharge $+1/2$:
\begin{equation}\label{eq:2hdmvisible}
\mathcal V_{\rm{visible}}^{\rm{2hdm}}=\mu_1^2|H_1|^2+\lambda_1|H_1|^4+\lambda_3 |H_1|^2|H|^2+\lambda_4|H_1 H|^2+(b H_1 H-\frac{\lambda_5}{2}(H_1 H)^2+h.c.)\,,
\end{equation}
where $H$ is the SM Higgs doublet, and where we have imposed a discrete $\mathbb{Z}_2$ symmetry that is only softly broken by the terms proportional to $b$. 

Generically, both $H$ and $H_1$ will get a VEV: $\langle H\rangle\equiv v\sin\beta/\sqrt 2,~\langle H_1\rangle\equiv v\cos\beta/\sqrt 2$. After EWSB, it is convenient to write the Lagrangian in the so-called Higgs basis \cite{Donoghue:1978cj}, where only one Higgs doublet gets a VEV: $\langle \phi_v\rangle=v/\sqrt 2,~\langle \phi_H\rangle=0$:
	\begin{equation}\label{eq:higgsbasis}
	\Phi_v =  \begin{pmatrix} G^+ \\\frac{1}{\sqrt{2}}( v + S_1 + i G^0) \end{pmatrix} ~,
	\qquad \qquad
	\Phi_H = \begin{pmatrix}  H^+\\\frac{1}{\sqrt{2}} (S_2 + i S_3)  \end{pmatrix}~.
	\end{equation}
In the limit $\lambda_i v^2\ll b$, we can write the mixing angle between these two fields $S_1$ and $S_2$ as\footnote{In the standard 2HDM language, this angle corresponds to the $\alpha-\beta+\pi/2$ combination.}  
\begin{equation}
\gamma\simeq F(\beta)\cdot \frac{\lambda v^2}{m_A^2}\ ,
\end{equation}
where $\lambda$ is a typical size of the quartic couplings of the potential, $F(\beta)$ is a function which encodes the $\tan\beta$ dependence, and $m_A$ is the mass of the pseudoscalar $S_3$: $m_A^2\simeq \frac{2b}{\sin2\beta}$. This scale controls the scale of the two heavy Higgses up to EW corrections. 

For $\lambda_i v^2\ll b$, the SM Higgs is given by $h\sim S_1+\gamma S_2$ and the new scalar by $S\sim S_2-\gamma S_1$. This mixing leads to the suppression of the SM Higgs coupling to massive gauge bosons as in Eq. (\ref{eq:SMHiggs}). 

Contrary to the singlet case, we can write explicit Yukawa terms for the second Higgs doublet, $H_1$. This adds an additional ($\tan\beta$) parametric dependence of the Higgs couplings to SM fermions and gauge bosons. Particularly, if we focus on a Type-I Yukawa structure, the Higgs couplings to fermions are 
\begin{align}
&\frac{g_{hff}}{g_{hff}^{SM}}= 1+\frac{\sin\gamma}{\tan\beta}+\mathcal O(\gamma^2)\,, \label{eq:SMHiggs2hdm}\\
&\frac{g_{Sff}}{g_{hff}^{SM}}= -\frac{1}{\tan\beta}+\sin\gamma+\mathcal O(\gamma^2)\,. \label{eq:SMHiggs2hdm}
\end{align}

The cross sections of the different production mechanisms of the second Higgs will scale in different ways. In a Type-I:
\begin{align}
&\sigma_S^{{\rm{VBF,VS}}}=\sin^2\gamma\cdot\sigma_h^{{\rm{VBF,VS}}}(m_A)\,,\\
&\sigma_S^{{\rm{ffS,ggS}}}=\frac{\sigma_h^{{\rm{ffS,ggS}}}(m_A)}{\tan^2\beta}+\mathcal O(\gamma^2)\,,
\end{align}
with $\sigma_h^{{\rm{VBF,VS,ffS,ggS}}}(m_A)$ the corresponding cross sections of a SM Higgs with mass $m_A$. Correspondingly, the scaling of the several branching ratios will differentiate between massive gauge bosons and fermions. 

Similarly to the singlet case, in the minimal setup presented in (\ref{eq:2hdmvisible}), having a displaced $S$ signature generically implies a reduction of the several $S$ production cross sections (either because of the reduction of the $S$ mixing with $h$, $\gamma$, or because of the reduction of the $S$ couplings with fermions). This tension is again easily solved by adding new interactions of the $H_1$ with a generic (displaced) dark sector of the type $-\lambda |H_1|^2X^2$. After EWSB, these interactions generate a coupling of $S$ with the dark sector that is independent on its mixing with the SM Higgs and on its couplings with fermions. As in the singlet model, the totality of decays of the doublet might involve prompt/visible final states; displaced final states; and invisible final states.

In conclusion, for the purposes of our study, the dark sector-enriched 2HDM can be parametrized by seven free parameters:

\begin{equation}
m_S,~\gamma,~\tan\beta,~\Gamma_{\rm{displaced}},~\Gamma_{\rm{invisible}},~\Gamma_{S\to hh},~c\tau\,,\end{equation}
where $c\tau$ is the life time of the dark sector particle $X$. 

There are ultimately several key differences between the singlet and doublet models. In the former case the relative ratio of heavy Higgs couplings to fermions and vectors is fixed, whereas in the latter case they are allowed to vary, potentially lessening the impact of powerful searches for decays into $ZZ$ and $hh$ without reducing the production cross section of the new states. Moreover, the complementarity between direct searches and Higgs coupling measurements changes significantly. In the alignment limit of the doublet model, $\gamma \approx 0$, the couplings of the observed 125 GeV Higgs are exactly as predicted by the Standard Model, while the heavy Higgs states retain nonvanishing couplings to Standard Model fermions whose strength depends on the value of $\tan \beta$. In this same limit, the heavy Higgses do not couple directly to Standard Model vectors, leaving only gluon fusion and $t \bar t S$ associated production as the dominant production modes, and the challenging $S \rightarrow t \bar t$ decay as the primary prompt decay mode (for the prospects for this decay mode see e.g. the recent studies \cite{Alvarez:2016nrz,Carena:2016npr,Czakon:2016vfr,Hespel:2016qaf,Craig:2016ygr,Gori:2016zto,Craig:2015jba,Jung:2015gta}). In this limit, LLP searches provide the main handle on the additional Higgs states.

\begin{figure}[h!]
\centering
\includegraphics[width=.49\textwidth]{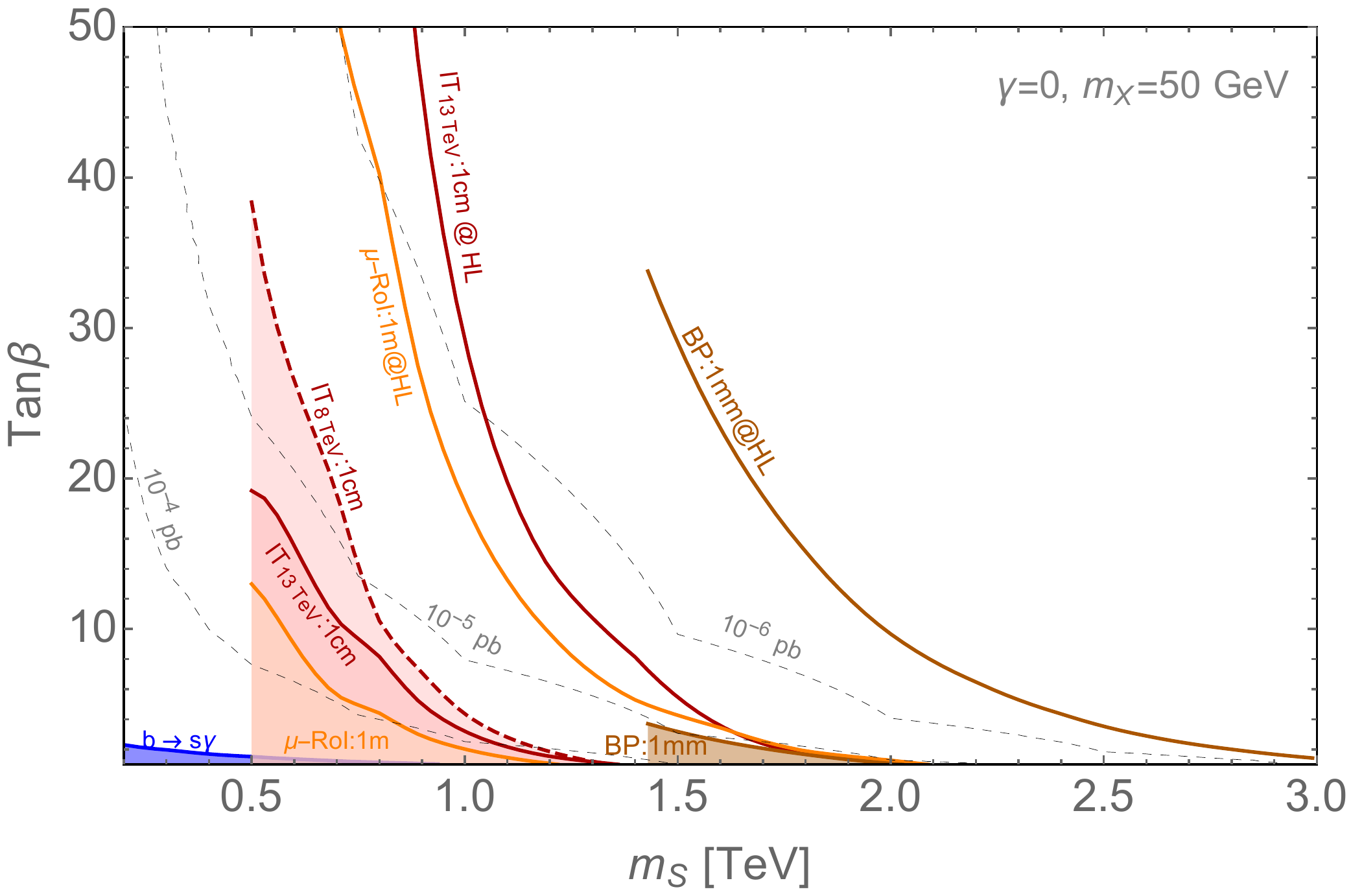}
\includegraphics[width=.49\textwidth]{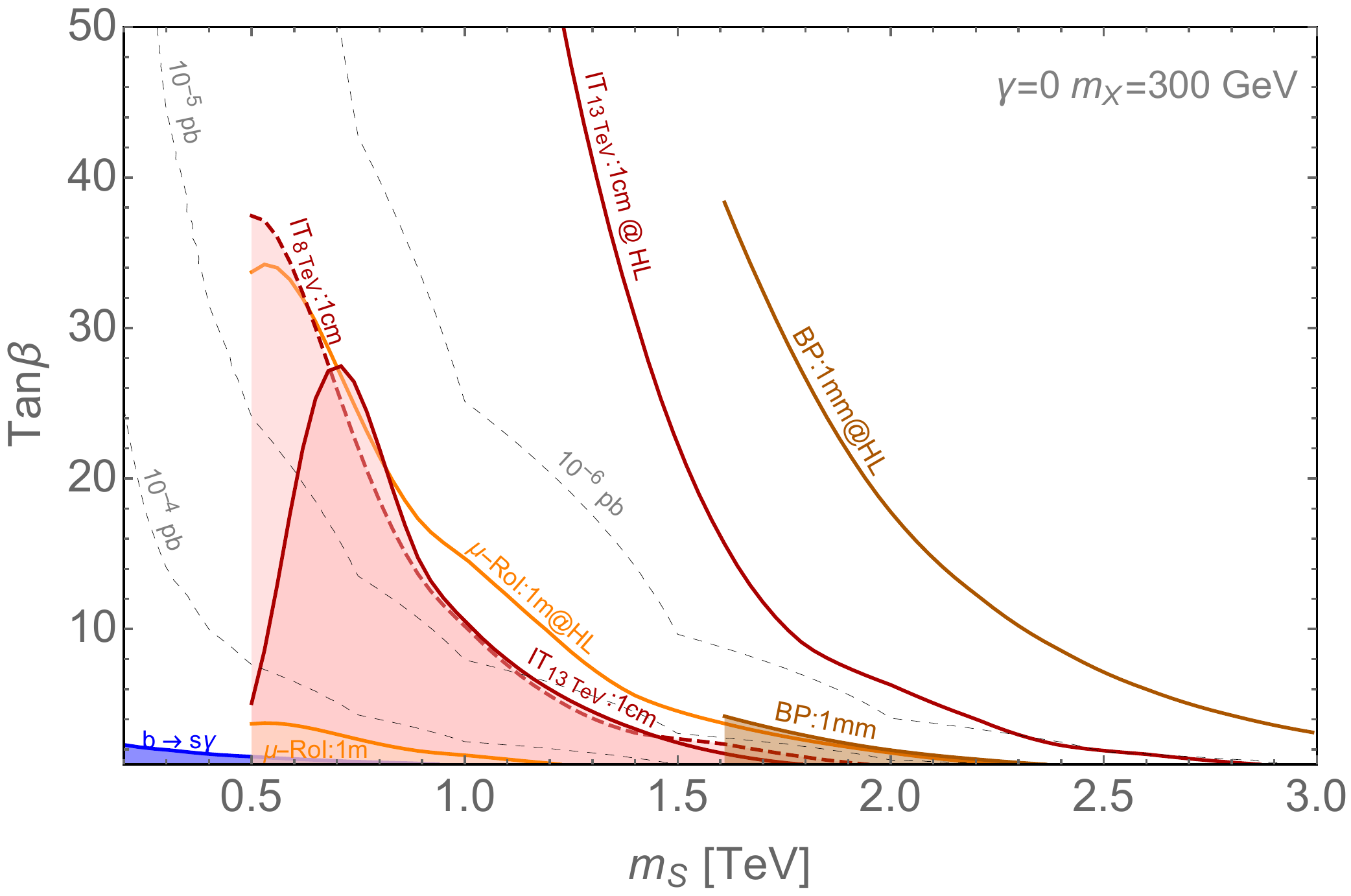}
\caption{Parameter space of the doublet Higgs scenario as a function of $m_S$ and $\tan\beta$ overlaid with current and projected constraints from direct searches. For the plot, we assume the alignment limit $\gamma=0$, the branching ratio into invisible equal to 0 and equal branching ratios into displaced objects and visible final states (mainly $t\bar t$). On the {\bf left/right} panel we show results for $m_X=50~{\rm{GeV}}/m_X=300$ GeV. The {\bf shaded regions} and {\bf solid lines} are as in Fig. \ref{fig:singlet}.
{\bf Gray contours} represent the rate for $t\bar tS, S\to XX$ (2 displaced). The blue shaded region on the bottom left of the plots is the region probed by the measurement of $b\to s\gamma$, under the assumption $m_S=m_{H^\pm}$.
\label{fig:summarydoublet}
}
\end{figure}

In Fig. \ref{fig:summarydoublet}, we show the reach of the several LHC displaced searches performed so far in the alignment limit $\gamma = 0$ (for which prompt di-boson and di-Higgs searches are ineffectual). The several dashed contours correspond to the cross section for a new signature that could be looked for at the LHC in the coming years: $t\bar tS, S\to XX$ (2 displaced). For a matter of simplicity we fix $\Gamma_{\rm{displaced}}=\Gamma_{\rm{visible}}$ and $\Gamma_{\rm{invisible}} = 0$, where in $\Gamma_{\rm{visible}}$ we include all SM decays of the $S$ boson \footnote{Additional (relatively weak) constraints on the parameter space of Fig. \ref{fig:summarydoublet} can arise from the measurement of flavor transitions. For example the measurement of the $b\to s \gamma$ transition leads to a bound at low values of $\tan\beta$ and light charged Higgs masses in Type-I 2HDMs: $m_{H^\pm}\gtrsim 250$ GeV (550 GeV) for $\tan\beta=2~ (1.5)$ \cite{Arbey:2017gmh}. See the blue region in Fig. \ref{fig:summarydoublet}.}.

\section{Neutral Naturalness}\label{sec:twinhiggs}

The simplified models presented in the previous sections capture the salient features of a wide variety of compelling scenarios for physics beyond the Standard Model. Particularly notable among these scenarios are approaches to the hierarchy problem such as Neutral Naturalness, which provide a motivated target for LLP searches at the LHC. Successfully addressing the hierarchy problem in these models requires:
\begin{enumerate}
\item[(1)] a hidden sector with a QCD-like gauge group whose confinement scale is close to that of the Standard Model. 
\item[(2)] one or more additional Higgs bosons that mix with the Standard Model Higgs doublet and couple to states in the hidden sector.
\end{enumerate}
These ingredients automatically lead to exotic decays of additional Higgs bosons, most notably decays to bound states of the hidden sector. Whether the decay products are long-lived on collider scales varies from model to model. In the mirror Twin Higgs \cite{Chacko:2005pe}, the lightest hidden sector bound states are typically pions that decay to lighter states in the hidden sector or, if such decays are unavailable, decay back to the Standard Model sufficiently slowly so as to be effectively detector-stable. In variants of the Twin Higgs such as the fraternal Twin Higgs \cite{Craig:2015pha}, however, there are not necessarily light quarks in the hidden sector, and the lightest hidden sector bound states are glueballs or bottomonia. Bound states with the same spin and angular momentum quantum numbers as the Higgs then typically decay back to the Standard Model on collider length scales by mixing with the Higgs, giving rise to classic displaced signatures typical of a Higgs portal Hidden Valley \cite{Strassler:2006ri}. Similar signatures arise in related avatars of neutral naturalness such as the Hyperbolic Higgs \cite{Cohen:2018mgv}, where a heavy singlet-like hyperbolic Higgs can decay into pairs of scalar top partners charged under a QCD-like hidden sector gauge group. The subsequent decays of these top partners into hidden sector glueballs lead to signatures analogous to that of the fraternal Twin Higgs.

Long-lived particles arising in incarnations of neutral naturalness are produced in decays of both the heavy Higgs(ses) and the SM-like Higgs at 125 GeV. While production of LLPs in exotic decays of the SM-like Higgs is a motivated target, reaching the irreducible branching ratio of $\sim 10^{-4}-10^{-5}$ across the full range of possible lifetimes will be difficult at the LHC due to limitations from trigger thresholds (see Refs~\cite{Csaki:2015fba, Curtin:2015fna}). Here we present a potentially more promising strategy which is to look for the production of LLPs from heavy Higgs decays, for which triggering is less of a limitation. While the partial widths for the heavy Higgs to hidden sector glueballs or bottomonia may not individually be large, the totality of heavy Higgs decay modes into the hidden sector, followed by cascade decays and/or annihilation into glueballs or bottomonia, leads to a large rate of LLP production in aggregate.

The essential properties of Twin Higgs models and their variants are summarized in e.g.~\cite{Burdman:2014zta}, to which we refer the reader for further details. In what follows, we will focus on the features of Twin Higgs models that are most relevant to the decays of the heavy Twin Higgs, exclusively considering the Fraternal Twin Higgs scenario \cite{Craig:2015pha} with the lightest glueball being the lightest dark particle (LDP) as a benchmark for displaced signatures. In order to make the paper self contained, more details about the structure of the dark sector are given in Appendix~\ref{app:fraternal}.

\subsection{Displaced decays of a Twin Higgs}
In Twin Higgs models the SM-like Higgs is a pseudo-Goldstone boson (PGB) of an approximate global $SU(4)$ symmetry spontaneously broken down to $SU(3)$, explaining the lightness of the Higgs with respect to the scale of new physics. The approximate $SU(4)$ symmetry arises from the potential of the two Higgs doublets, a doublet, $H_A$, charged under the Standard Model and a second doublet, $H_B$, charged under a mirror copy thereof. The two copies of the Standard Model are related by a $\mathbb{Z}_2$ symmetry under which $A \leftrightarrow B$. The most general renormalizable Higgs potential for the visible ($H_A$) and the Twin Higgs ($H_B$) doublets is given by
\begin{equation}
V=\lambda\left(\vert H_A\vert^2+\vert H_B\vert^2\right)^2-m^2\left(\vert H_A\vert^2+\vert H_B\vert^2\right)+\kappa\left(\vert H_A\vert^4+\vert H_B\vert^4\right)+\tilde{\mu}^2\vert H_A\vert^2+\rho \vert H_A\vert^4,
\end{equation}
where $\lambda$ and $m^2 (>0)$ are the $SU(4)$ preserving terms, $\kappa$ preserves the $\mathbb{Z}_2$ mirror symmetry that exchanges $A\leftrightarrow B$, but breaks $SU(4)$, and $\tilde\mu$ and $\rho$ are the $\mathbb{Z}_2$ breaking terms.

We can parametrize the Higgs VEVs as $\langle H_{A,B}\rangle=v_{A,B}/\sqrt 2$. For the ease of notation we define  
\be
f^2\equiv v_A^2+v_B^2\approx  \frac{m^2}{\lambda},~~~
\sigma\equiv\frac{\lambda\tilde{\mu}^2}{m^2}\approx\frac{\tilde{\mu}^2}{f^2},
 \ee
where the approximated expressions hold in the $\kappa,\rho,\sigma\ll\lambda$ limit. In the same limit, after EWSB, the mass of the twin Higgs and the mixing angle are given by
\begin{equation}
m_{\phi}^2\approx2\lambda f^2\,,~~\\
\sin\theta\approx \frac{v}{f}\label{eq:mix}.
\end{equation}
In addition, the trilinear coupling of the Twin Higgs to two SM-like Higgses is given by
\begin{equation}
A_{\phi hh}\equiv \partial_{\phi}\partial_h^2 V\vert_{h,\phi=0}\approx \frac{m_{\phi}^2}{ f}\label{eq:trilinearhT}\,.
\end{equation}
 In general, there can be substantial deviations from these approximate expressions, especially in supersymmetric UV completions of Twin Higgs models where typically $\lambda\lesssim 0.5$ and $\kappa/\lambda\gtrsim 1/5-1/3$ \cite{Craig:2013fga,Katz:2016wtw}. Formulas at all orders can be found in \cite{Katz:2016wtw} and will be used in the plots that follow. 


The tree-level interactions of the two Higgs bosons with the SM and Twin gauge bosons are 
\begin{align}
{\cal L} & = \left(  \frac{g}{2 c_W} M_Z  Z^\mu Z_\mu + g M_W W^{+ \mu} W_\mu^- \right) \left(  h \cos\theta + \phi\sin\theta\right) \nonumber \\
& + \left(  \frac{g}{2 c_W} M_{Z_B}  Z^\mu_B Z_{B \mu} + g M_{W_B} W^{+ \mu}_B W_{B \mu^-} \right) \left(  -h \sin\theta + \phi\cos\theta  \right)\,,
\end{align}
where the masses of the twin gauge bosons, $V_B$, are related to the masses of the corresponding SM gauge bosons, $V$, by $m_{V_B}= m_Vf/v$.

In the mirror Twin Higgs, the fermion content and Yukawa couplings are mirror copies of the Standard Model. However, the majority of these states are inessential to the stabilization of the weak scale, and treating the masses and couplings of irrelevant states as free parameters leads to the more minimal Fraternal Twin Higgs \cite{Craig:2015pha}. In the Fraternal Twin Higgs, the relevant Yukawa interactions are those of the third generation quarks and leptons
\begin{equation}
\mathcal{L}_{\text{Yuk}}\!=\!y_t\left(H_A t_A^{l} t_A^{r}+ H_B t_B^{l} t_B^{r}\right)+y_b \left(H_A b_A^{l} b_A^{r}+\delta y_b H_B b_B^{l} b_B^{r}\right)+y_\tau  \left(H_A \tau_A^{l} \tau_A^{r}+\delta y_\tau H_B \tau_B^{l} \tau_B^{r}\right)\label{eq:yuk},
\end{equation}
where we have fixed $\delta y_t= 1$ since naturalness requires the top and twin-top Yukawa couplings to be essentially identical close to the cutoff. These couplings define the dominant phenomenology of the Twin Higgs.

\begin{figure}[h!]
\centering
\includegraphics[width=.8\textwidth]{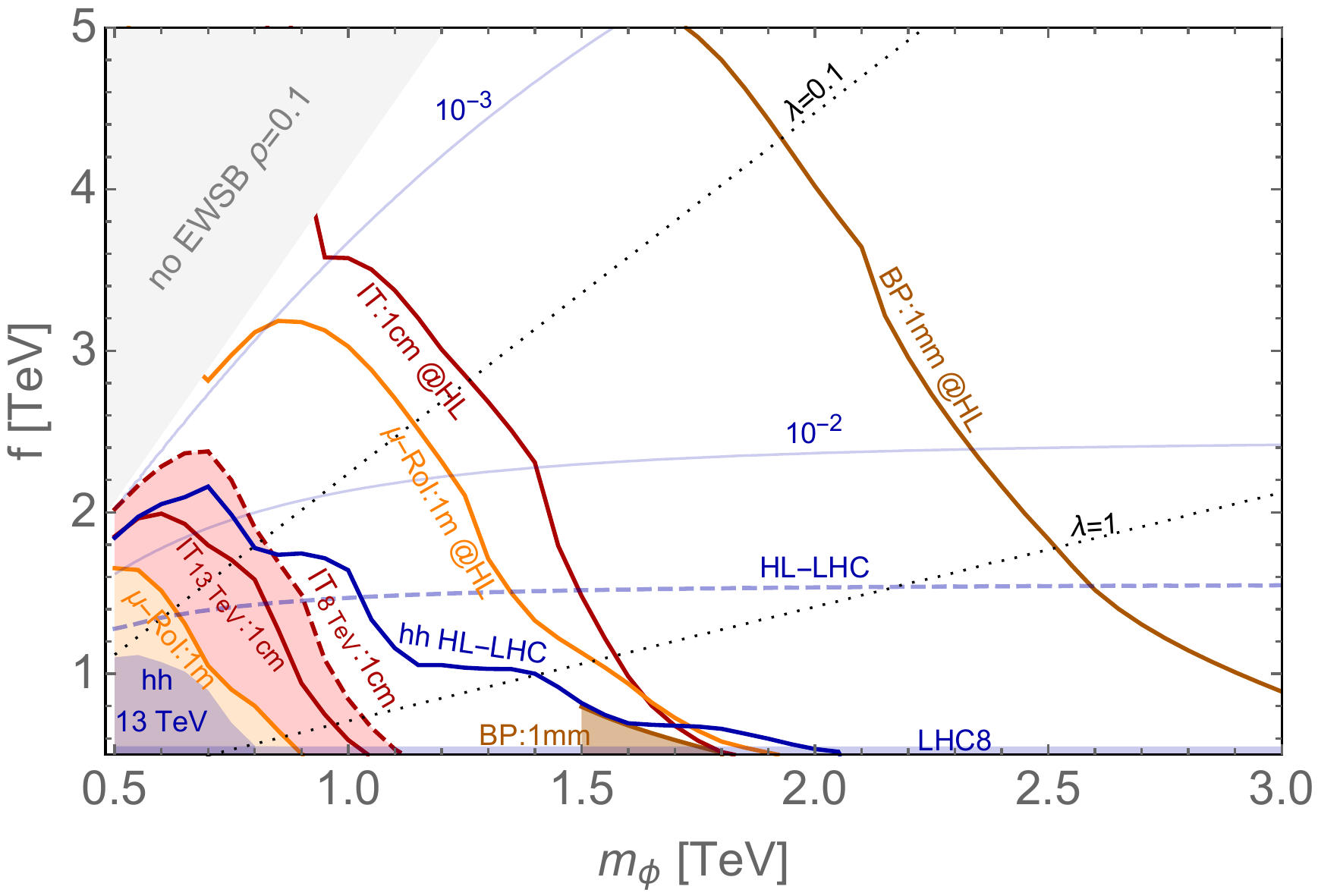}
\caption{Parameter space of the Fraternal Twin Higgs model as a function of the Twin Higgs mass, $m_\phi$, and $f$ overlaid with current and projected constraints from direct searches. We vary $\delta y_b$ and $\delta g$ in order to fix the lightest glueball mass to 50 GeV across the entire parameter space and simultaneously single out representative values of $c \tau$ (see text for details). We assume that 100\% of the $t_B \bar t_B$ and $Z_B$ $Z_B$ pairs from the decays of the Twin Higgs eventually decay into glueballs. {\bf Solid blue lines} indicate the value of $\sin^2 \gamma$, while the {\bf dashed black lines} correspond to constant values of the $SU(4)$-symmetric quartic, $\lambda$. The {\bf gray shaded} region cannot produce EWSB for $\rho=0.1$. The {\bf orange shaded} region indicates the exclusion from the 13 TeV ATLAS muon RoI search for $c \tau=1\text{ m}$ for the lightest glueball. The assumptions about the glueball production modes are detailed in the text. The {\bf red shaded} indicates the exclusion for the CMS IT search at 13 TeV with $c \tau=1\text{cm}$. For the same decay length, we show as a {\bf dashed red line} the extrapolation of the 8 TeV search. The {\bf dark orange shaded} region shows the exclusion from the CMS BP search at 13 TeV for $c\tau=1\text{ mm}$. The {\bf orange, red and dark orange lines} are projections of the displaced searches to HL-LHC under the assumption of zero background. The small {\bf blue shaded} region on the lower left corner of the figure indicates the exclusion from resonance di-Higgs production, which is stronger than the $ZZ$ searches in this case. The {\bf blue line} shows the projections at HL-LHC. The {\bf blue shaded horizontal band} at low values of $f$ is probed by 8 TeV Higgs coupling measurements, while the {\bf light blue dashed line} shows the corresponding extrapolation to the HL-LHC.
\label{fig:twinmoney}
}
\end{figure}

In Fig. \ref{fig:twinmoney}, we show the status of a representative slicing of parameter space of the Fraternal Twin Higgs model. We refer the reader to Appendix \ref{sec:ratesTwin} for details on the calculation of the Twin Higgs rates into visible and displaced final states. While the cascade decays of the Twin Higgs into glueballs are quite complex and require a detailed treatment of dark showering to capture correctly, we take the glueball final states to be well-described by our LLP pair-production simplified model for the purposes of illustrating the potential reach of LLP searches. For each point in the figure, we choose $\delta y_b$ and $\delta g$ to fix the mass of the lightest glueball to 50 GeV, and also to single out specific values of $c \tau$ for the lightest glueball that highlight the sensitivity of various LLP searches. We have chosen $\rho = 0.1$ for this figure, which leads to a broader parameter space with successful EWSB compared to $\rho = 0$. One consequence of this choice is that the rate for $\phi \rightarrow hh$ is enhanced compared to the case of $\rho = 0$, so that limits from prompt decays in current data and HL-LHC projections are driven by $\phi \rightarrow hh \rightarrow 4b$ (blue shaded region in the figure). In some sense this gives the strongest possible projection for the reach of visible searches, making the potential effectiveness of LLP searches all the more apparent. The 13 TeV ATLAS muon RoI search, the CMS IT and BP searches at 13 TeV suggest that LLP searches with current 13 TeV LHC data have the potential to provide broad coverage of the parameter space for Twin Higgs masses up to $\sim 1.5$ TeV, at least for the representative parameters chosen here. Suitable searches at the HL-LHC could potentially extend coverage to masses of order $\sim 2.5$ TeV, significantly exceeding the reach of searches for prompt decay products of the Twin Higgs. Of course, we emphasize that we have shown only a particular slicing of the Twin Higgs parameter space to illustrate the value of LLP searches, and the coverage of direct and displaced searches is quite sensitive to varying the lightest glueball mass and lifetime.

\subsection{New displaced signals from Twin Supersymmetry}

Ultimately, the Twin Higgs and its relatives are only a solution to the ``little'' hierarchy problem, insofar as they do not stabilize the scale $f$ itself against sensitivity to much higher scales. This necessarily entails the UV completion of Twin Higgs models into solutions to the ``big'' hierarchy problem such as supersymmetry or compositeness. Supersymmetric UV completions of the Twin Higgs \cite{Falkowski:2006qq, Chang:2006ra, Craig:2013fga, Katz:2016wtw} are particularly compelling, as they automatically explain the approximate $SU(4)$ symmetry of the Higgs potential and are in better agreement with precision electroweak constraint. Such UV completions necessarily predict the further extension of both the Standard Model Higgs sector and the Twin Higgs sector into two-Higgs-doublet models, significantly increasing the number of Higgses with potentially large branching ratios into LLPs.

As discussed in \cite{Katz:2016wtw}, the MSSM-like Higgses can be lighter than the Twin Higgs, in which case the doublet Higgs simplified model presented in Section \ref{sec:doublet} becomes a more appropriate characterization of the relevant phenomenology.\footnote{The two Higgs doublet model introduced in SUSY completions of the Twin Higgs is necessarily of Type II, for which the scaling of fermion couplings differs relative to the doublet model presented in Section \ref{sec:doublet}.} This has several novel implications. First, the relative branching ratios of the heavy Higgses into LLPs, Standard Model vectors, and Standard Model fermions changes significantly from the singlet case, potentially weakening bounds in prompt decay modes. Second, fermionic associated production modes such as $b \bar b$ and $t \bar t$ associated production can become more significant, providing a new handle for LLP searches. Third, there is a new massive CP-odd Higgs arising from the additional doublet charged under the Standard Model. This opens a new {\it CP-odd} Higgs portal to the hidden sector, corresponding to new dimension-5 and dimension-6 operators that allow the decay of CP-odd bound states. In particular, upon integrating out the heavy radial mode, the theory contains effective interactions of the form
\begin{equation}
\mathcal{L}_{\text{MSSM}}^{\text{A-portal}}=c^{gg}_{ud}\frac{h_u^A h_d^A}{\Lambda^2}\tilde{G}_{B}\cdot G_B+ c^{bb}_{ud}\frac{h_u^A h_d^A}{\Lambda}\bar{b}_B\gamma_5 b_B+\dots
\end{equation}
which lead, among other things, to mixing between the pseudoscalar Higgs, $A$, and both the $0^{-+}$ glueball and the CP-odd bottomonium $\chi_0^{-+}$. The latter is typically the lightest of the bottomonia states, which might otherwise be detector-stable in the fraternal Twin Higgs. The lifetimes of these CP-odd states are necessarily longer than that of the CP-even states, since the partial widths for their decays into the Standard Model are suppressed by a relative factor of $m_h^4/m_A^4$, but this nonetheless opens up a rich variety of new long-lived channels connecting the Standard Model and the dark sector.

\section{Conclusion}\label{sec:conclusions}

The search for additional Higgs bosons is a crucial component of the physics program at the Large Hadron Collider. While the Higgs sector of Nature may trivially contain only the observed 125 GeV Higgs boson, there could readily be a variety of states associated with electroweak symmetry breaking -- potentially motivated by resolutions to puzzles of the Standard Model, or merely another instance of plenitude in the spectrum of fundamental particles.

So far, the LHC has focused on searches for new Higgs bosons decaying to Standard Model (SM) particles, as obtained in many minimal beyond-the-Standard Model theories.
Considerably less attention has been devoted to possible exotic decays of additional Higgs bosons into new particles that undergo further decay into Standard Model (or potentially invisible) final states. These exotic decays may provide the main discovery channels for both the heavy Higgses and the new states alike. 
In many well-motivated BSM scenarios, the new states in question are long-lived. When produced at relatively high energies from the decay of the additional Higgses, these long-lived particles can lead to spectacular signatures at the LHC.

In this paper, we have initiated the systematic study of a second Higgs boson at ``the lifetime frontier''. We reinterpreted a variety of LHC searches for both prompt and displaced decays at 8 and 13 TeV in the framework of a heavy Higgs scalar. 
We then analyzed the impact of these limits on simplified models containing a new singlet scalar or an additional electroweak doublet scalar coupled to a pair of long lived particles. The set of parameters that fully characterize the phenomenology of these models is relatively limited, offering an interesting opportunity to study the complementarity of LHC direct searches for the new scalar (decaying either in conventional ways or to long lived particles) and LHC Higgs coupling measurements in probing the parameter space.

These simplified models can be mapped on to well motivated extensions of the SM. Twin Higgs models and their supersymmetric completions offer a particularly clean realization of these two scenarios. In Twin Higgs models, the additional Higgs bosons can decay to hidden sector bound states (e.g. glueballs) that decay into SM particles with potentially observable lifetimes. The sensitivity of current and projected searches for  displaced decays demonstrates that they can provide the strongest constraint on the parameter space of these theories. This can be achieved in spite of the fact that branching ratios into long-lived hidden sector bound states are typically subdominant compared to branching ratios into di-bosons. Finally, we have investigated the prospects for probing or discovering new Higgs bosons at the HL-LHC, highlighting a few new search strategies -- most notably searches for $t \bar t$ associated production of heavy Higgses with decays to LLPs, and the development of boosted strategies to access larger mass differences between the heavy Higgs and the LLP -- that could be adopted by the LHC collaborations to maximize the discovery potential.

\bigskip

\noindent {\bf Note Added:} While this work was under preparation, we became aware of \cite{Kilic:2018sew}, which investigates the prospects for discovering the radial mode of the fraternal Twin Higgs model in a specific search for long-lived particles.

\section*{Acknowledgments}
We thank Simon Knapen, Ted Kolberg, Zhen Liu, Simone Pagan Griso, and Brian Shuve for helpful conversations. We thank Eric Kuflik and Oren Slone for discussions about the reinterpretation procedure for the 8 TeV CMS LLP search used in \cite{Csaki:2015uza}. We thank Can Kilic and Chris Verhaaren for correspondence regarding \cite{Kilic:2018sew}. We thank an anonymous referee for a thoughtful report. The research of SAF, NC, and SK is supported in part by the US Department of Energy under the Early Career Award DE-SC0014129 and the Cottrell Scholar Program through the Research Corporation for Science Advancement. The research of SG is supported in part by the NSF CAREER grant PHY-1654502. The collaboration of NC and DR is supported in part by the Israeli-US Binational Science Foundation (BSF) Grant No.~2014230. We also thank the Kavli Institute of Theoretical Physics for hospitality during the completion of parts of this work, and corresponding support from the National Science Foundation under Grant No. NSF PHY-1748958.

\appendix 
\section{Displaced searches at 8 TeV}\label{app:8TeVsearches}
We present representative limits obtained from the reinterpretation of two displaced searches at $\sqrt{s} = 8$ TeV: the ATLAS search based on the muon Region of Interest trigger \cite{Aad:2015uaa} and the CMS search based on the Inner Tracker trigger \cite{CMS:2014wda}. Since both these searches have an updated version at 13 TeV discussed in Sec.~\ref{sec:13TeVsummary}, we will explicitly compare them with their updates at 13 TeV.  Our simulation procedure for obtaining the signal efficiency for the 8 TeV searches is summarized in the next appendix. 

\begin{figure}[t]
\includegraphics[width=.48\textwidth]{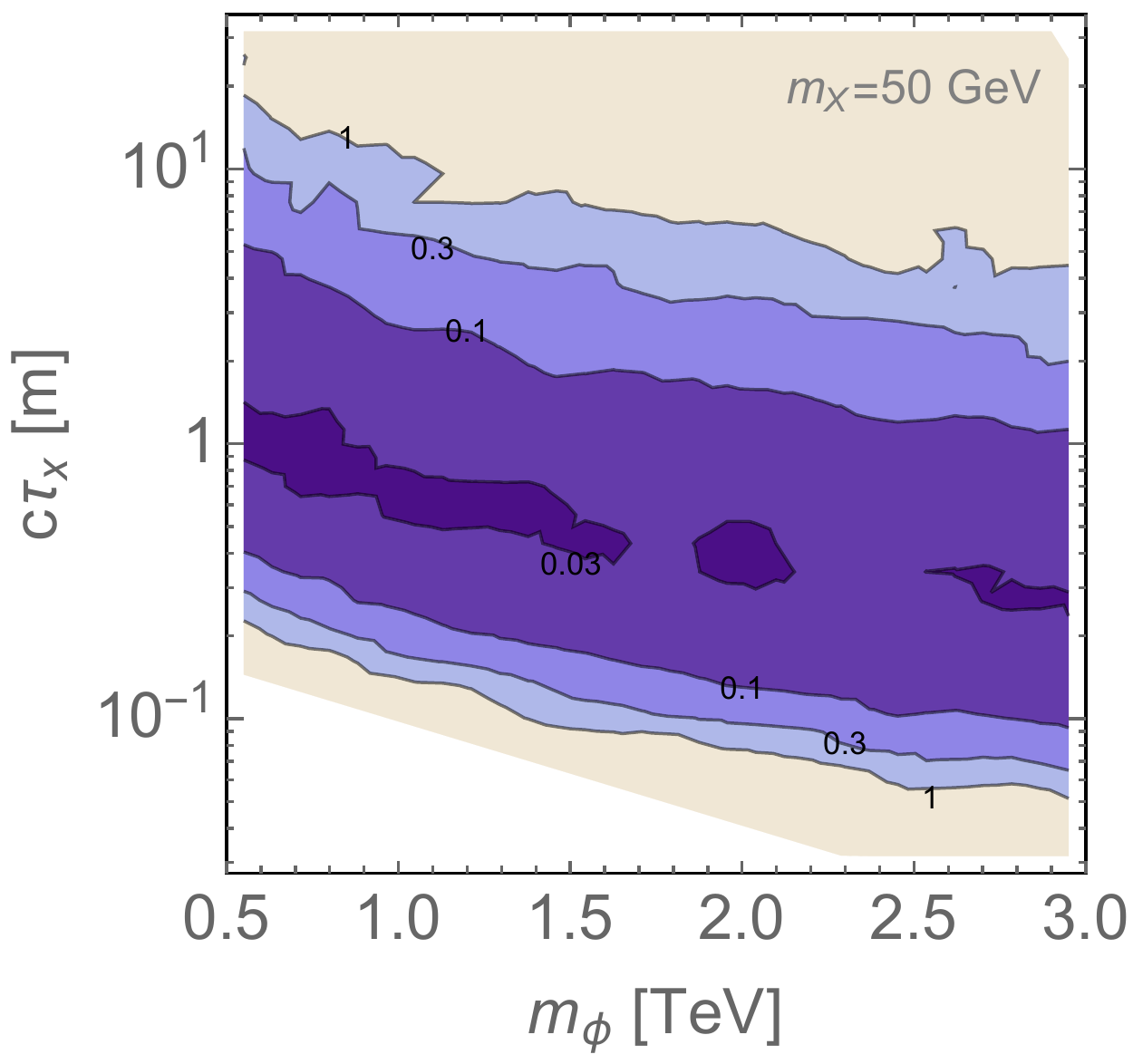}\hfill
\includegraphics[width=.48\textwidth]{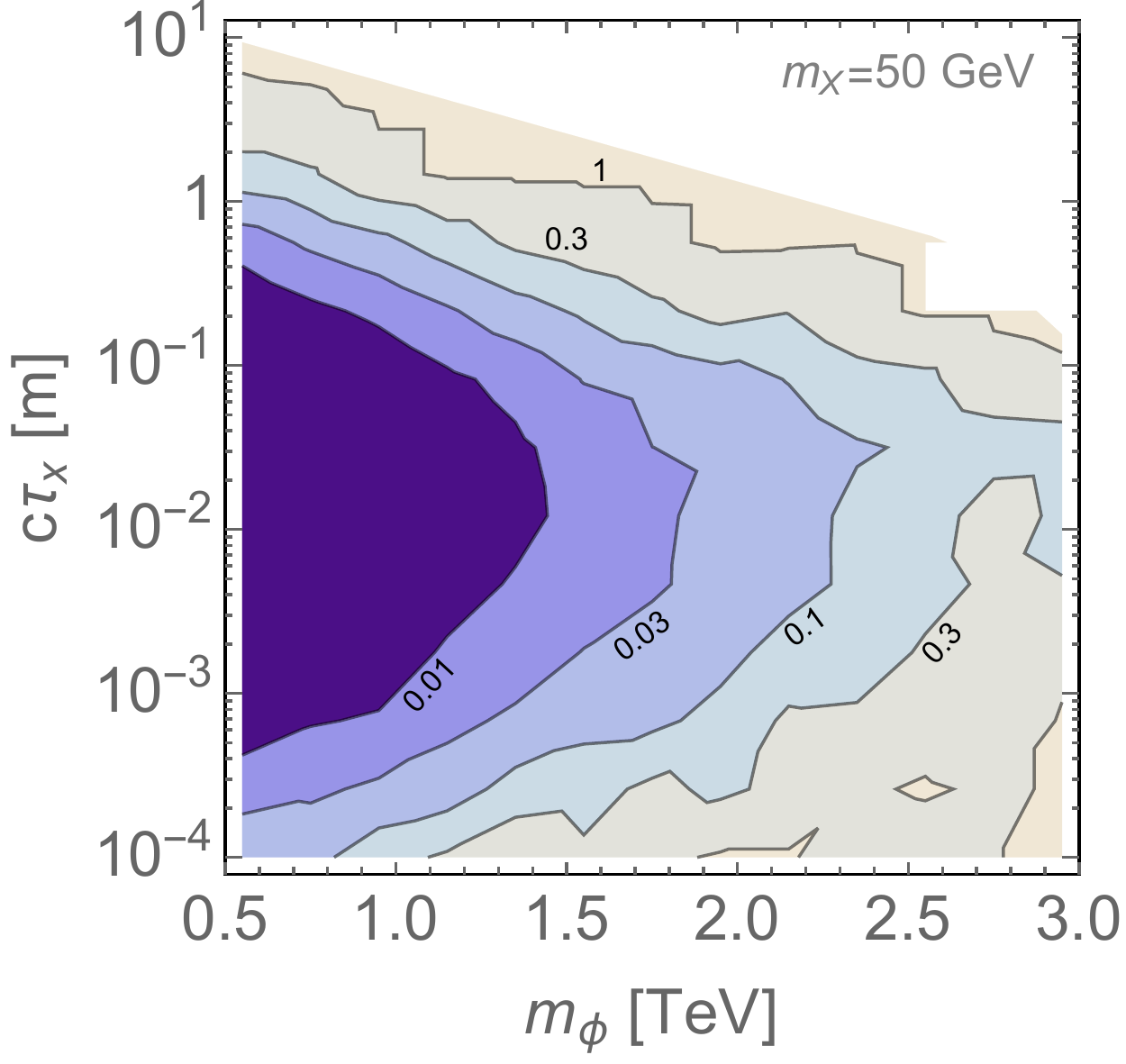}\\
\includegraphics[width=.48\textwidth]{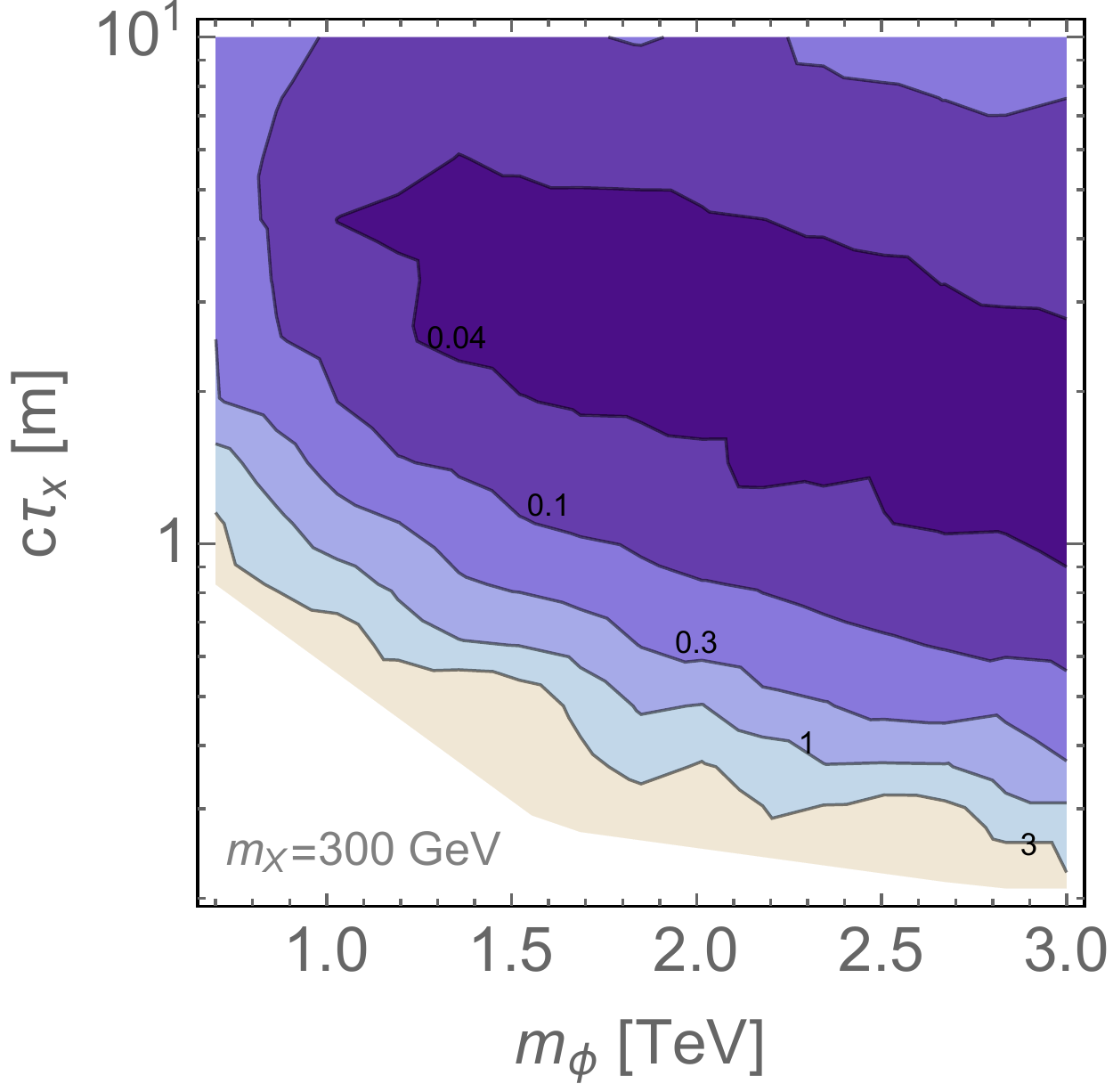}\hfill
\includegraphics[width=.48\textwidth]{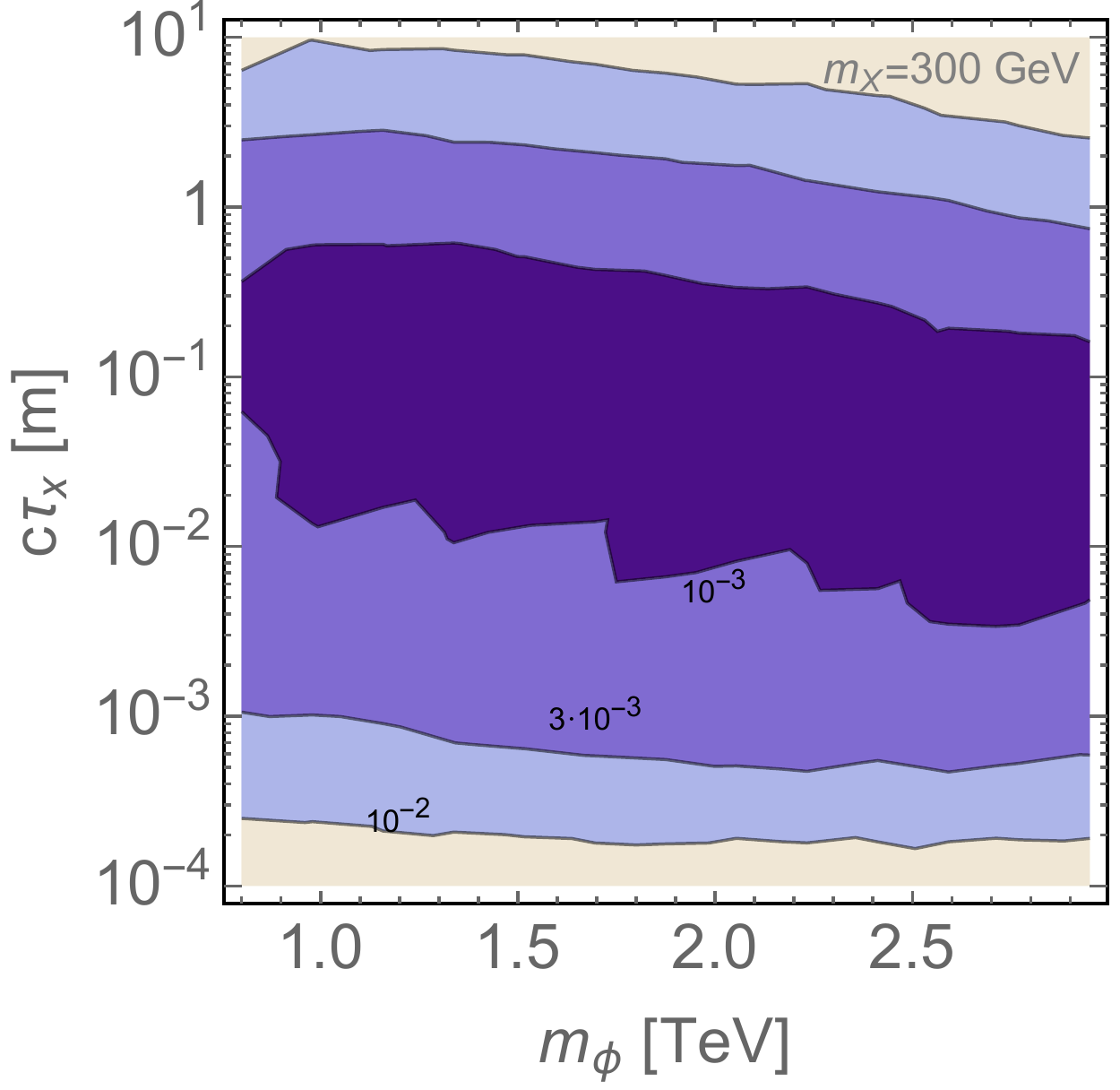}
\centering
\caption{Contours of the excluded signal strength $\sigma_{\phi}\cdot \text{BR}$ in the ($m_\phi$, $c\tau_X$) plane for the two LLP searches discussed in the text: {\bf Left:} ATLAS $\mu$-RoI search at 8 TeV \cite{Aad:2015uaa} {\bf Right:} CMS IT search at 8 TeV~\cite{CMS:2014wda}. The upper/lower panels refer to $m_X=50\text{ GeV}$/$m_X=300\text{ GeV}$, respectively.
}
\label{fig:8TeVLLP}
\end{figure} 

In Fig.~\ref{fig:8TeVLLP}, we show the excluded regions in the ($m_\phi$, $c\tau_X$) plane. The contours of the $\mu$-RoI search looks very similar to the ones obtained in Sec.~\ref{sec:13TeVsummary} for the 13 TeV analysis. Indeed, the 13 TeV exclusion can be roughly obtained by rescaling the 8 TeV one linearly with the luminosity $L_{8\text{ TeV}}/L_{13\text{ TeV}}\simeq 0.9$. The reason behind this simple rescaling is that very little changed in the data selection from the 8 TeV to the 13 TeV search.

The contours of the CMS IT search at 8 TeV are also qualitatively similar to the 13 TeV ones, but the actual value of the signal efficiency deviates significantly form the 13 TeV analysis. In particular, we already showed in Sec.~\ref{sec:13TeVsummary} that the 8 TeV data selection is more efficient in selecting the signal coming from a highly boosted daughter resonance, $X$. Indeed, while the 8 TeV and 13 TeV analyses have a qualitatively similar requirement that there are at most two ``prompt tracks'', the 8 TeV analysis does not require that there is at least one ``displaced track''. This feature makes the coverage of the 8 TeV search superior to the one of the 13 TeV one for $m_X=50\text{ GeV}$. 

In addition to these two searches, we note that in the ATLAS Ref.~\cite{Aad:2015uaa} a $j+\text{MET}$ trigger is used to tag shorter displacements in the inner tracker. Moreover, in Ref.~\cite{Aad:2015asa} a different search strategy is put forward for particles decaying to jets in the ATLAS hadronic calorimeter. We checked that both these analyses are less sensitive than a combination of the CMS inner tracker analysis \cite{CMS:2014wda} and the ATLAS RoI trigger search \cite{Aad:2015uaa} for a heavy resonance decaying into a pair of displaced daughter particles further decaying into hadronic jets at $\sqrt{s} = 8$ TeV.

\section{Reinterpretation of LHC searches}

For all our reinterpretations we use \texttt{FeynRules} \cite{Alloul:2013bka} to define signal models for our analyses and generate the signal events using \texttt{MadGraph 5} \cite{Alwall:2014hca}. Particle decays, hadronization, and showering are modeled with \texttt{Pythia 8} \cite{Sjostrand:2007gs}, and detector effects are taken into account using \texttt{Delphes} \cite{deFavereau:2013fsa}. 

\subsection{ATLAS Muon Regions of Interest Analyses} \label{app:ATLASmuon}

\begin{figure} 
  \centering
    \includegraphics[width=0.49\textwidth]{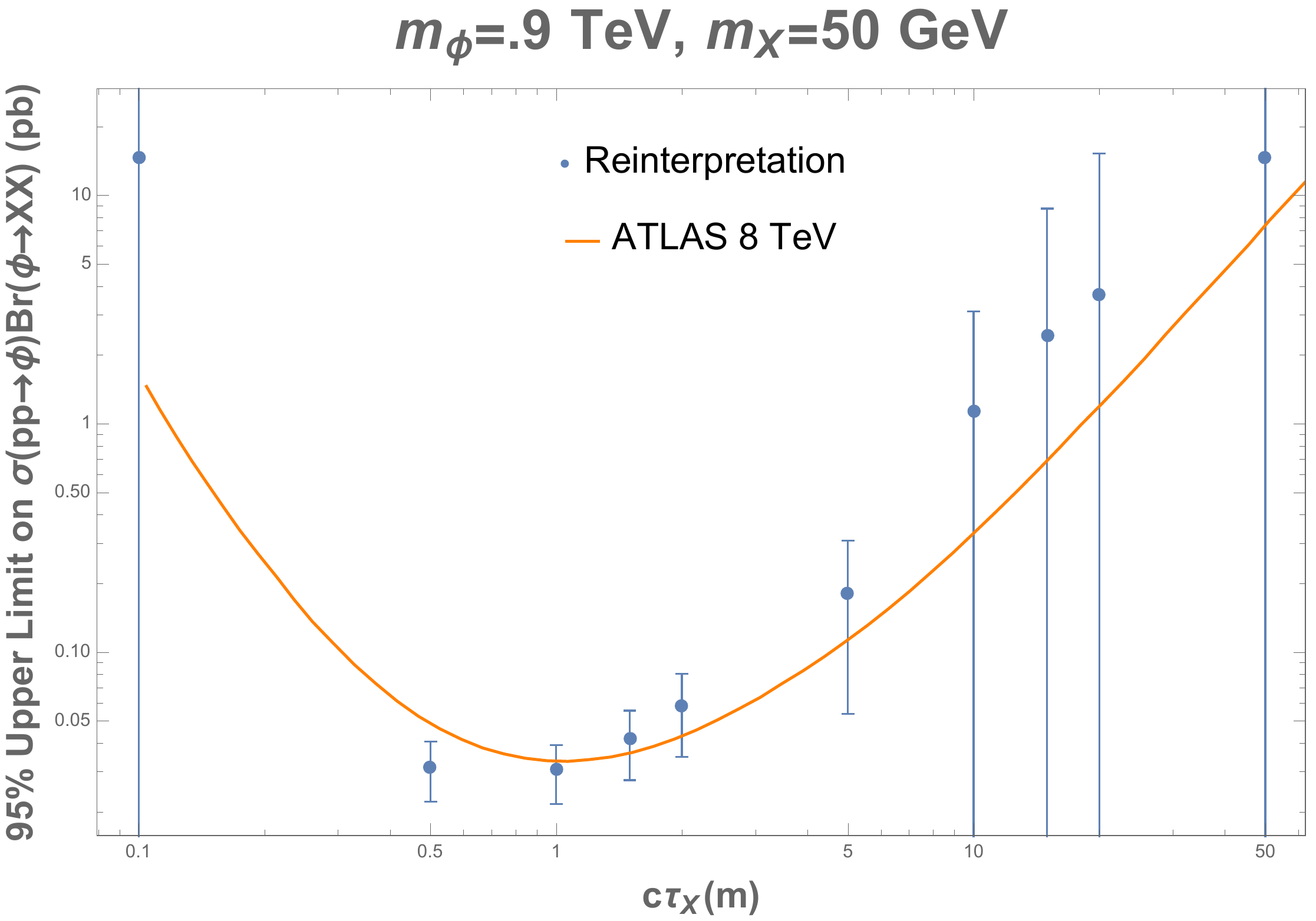}
    \includegraphics[width=0.49\textwidth]{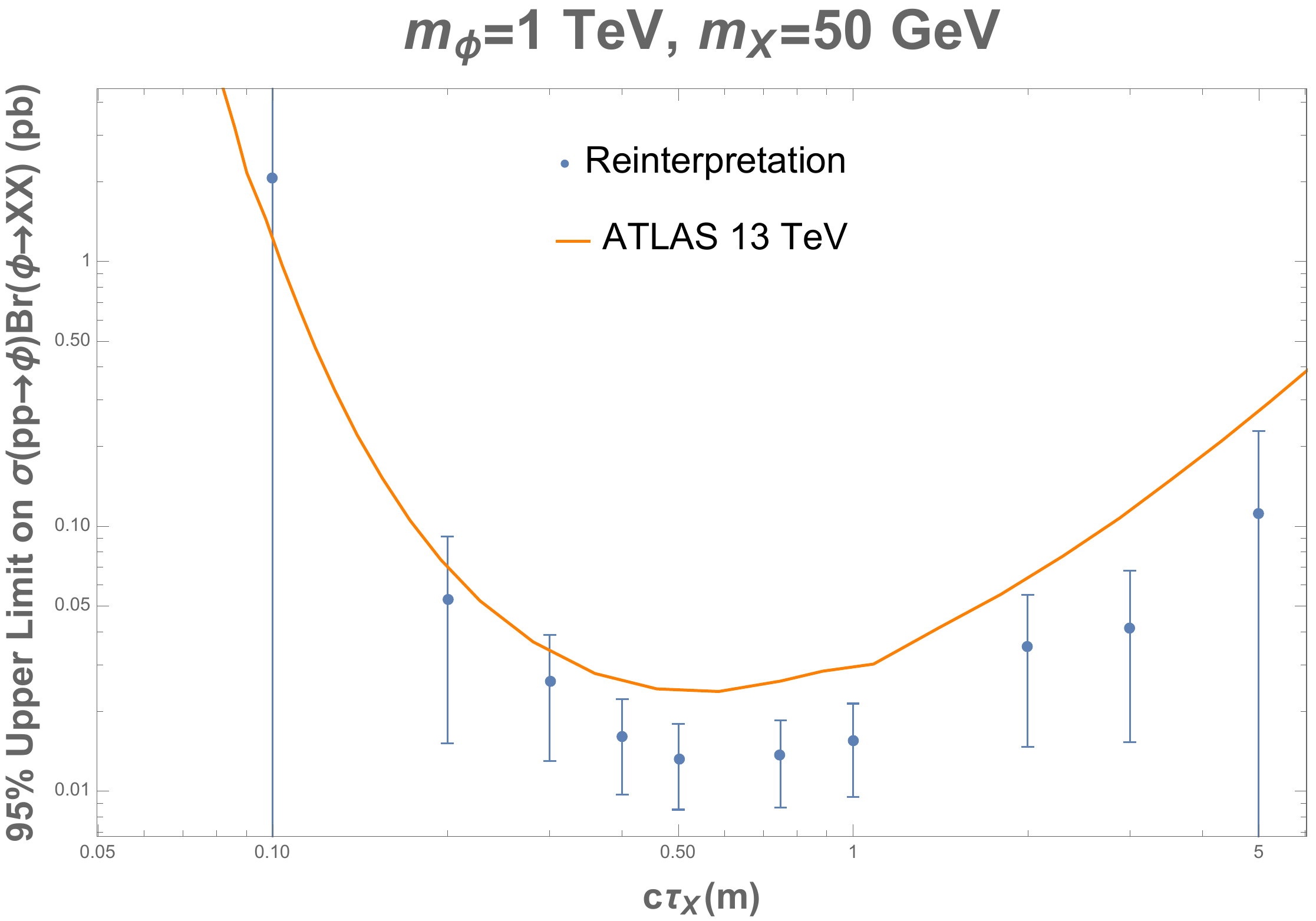}
  \caption{In both panels the {\bf orange line} is the result given by the experimental collaboration and the {\bf blue dots} correspond to the result of our reinterpretation, where the error band are estimated from our montecarlo by taking the variable we are plotting and multiplying it by $2/\sqrt{N}$ where $N$ are the number of events passing the cuts. {\bf Left:} A comparison between the $95\%$ sigma times branching ratio limits placed by the 8 TeV ATLAS analysis \cite{Aad:2015uaa} and our reinterpretation as described in Appendix \ref{app:ATLASmuon}, for an example mass point $m_\phi = 0.9$ TeV, $m_X = 50$ GeV. {\bf Right:} A comparison between the $95\%$ cross section times branching ratio limits placed by the 13 TeV ATLAS analysis \cite{Aaboud:2018aqj} and our reinterpretation as described in Appendix \ref{app:ATLASmuon}, for an example mass point $m_\phi = 1$ TeV, $m_X = 50$ GeV.} \label{fig:atlasroival}
\end{figure} 

As discussed in the previous section, the procedures for the ATLAS searches performed at $8$ TeV \cite{Aad:2015uaa} and at $13$ TeV \cite{Aaboud:2018aqj} do not differ greatly. We therefore implement similar analysis pipelines for the two searches, which differ primarily in the triggering and vertex reconstruction efficiencies, as informed by auxiliary material made available by ATLAS. For these reinterpretations we generate 5k $pp \rightarrow \phi \rightarrow XX (\rightarrow b\bar b b \bar b)$ signal events at each $(m_X, m_\phi, c\tau_X)$ signal point.

For both searches, we use a reinterpretation strategy inspired by the approach taken in \cite{Liu:2015bma} for the 7 TeV muon RoI analysis. For the reinterpretation of the 8 TeV search, we define `barrel' and `endcap' trigger regions with flat, non-zero efficiencies. LLP decays occurring at $3.5 \text{ m} < r < 7.0 \text{ m}$, $| \eta | < 1.0$ are given a $0.65$ efficiency for activating the Muon RoI Cluster Trigger and decays in $6.0 \text{ m} < z < 12.5 \text{ m}$, $1.0 < |\eta| < 2.5$ are given a $0.60$ efficiency. An event passes the trigger if it contains at least one LLP decay which activates the trigger. We then require the reconstruction of two displaced vertices within the regions $3.5 \text{ m} < r < 8.0 \text{ m}$, $| \eta | < 1.0$ and $5.5 \text{ m} < z < 13.5 \text{ m}$, $1.0 < |\eta| < 2.5$ which occurred at a maximum delay from light speed of $0.7$ ns. We give each barrel vertex a $0.30$ efficiency for being reconstructed and each endcap vertex a $0.60$ efficiency, independent of the triggering requirement. We furthermore approximate a n isolation requirement by vetoing events with LLP decay products of energy $> 15 \text{ GeV}$ flowing into the HCAL from the barrel.

For the reinterpretation of the 13 TeV search, we modify our trigger efficiencies to $0.60$ within $4.0 \text{ m} < r < 6.5 \text{ m}$, $| \eta | < 1.0$ and $6.0 \text{ m} < z < 12.0 \text{ m}$, $1.0 < |\eta| < 2.5$. For our active vertex reconstruction volumes we now take $4.0 \text{ m} < r < 7.0 \text{ m}$, $| \eta | < 1.0$ with an efficiency of $0.30$ and $6.0 \text{ m} < z < 12.0 \text{ m}$, $1.0 < |\eta| < 2.5$ with efficiency of $0.60$. No change is made to the time delay or isolation requirements.

We validate this reinterpretation by comparing our cross section $\times$ branching ratio limits to benchmarks given in the ATLAS analyses \cite{Aad:2015uaa}, \cite{Aaboud:2018aqj}, as can be seen in Figure \ref{fig:atlasroival}.

\subsection{CMS Inner Tracker Analysis at 8 TeV}\label{app:CMS8TeV}

\begin{figure} 
  \centering
    \includegraphics[width=0.49\textwidth]{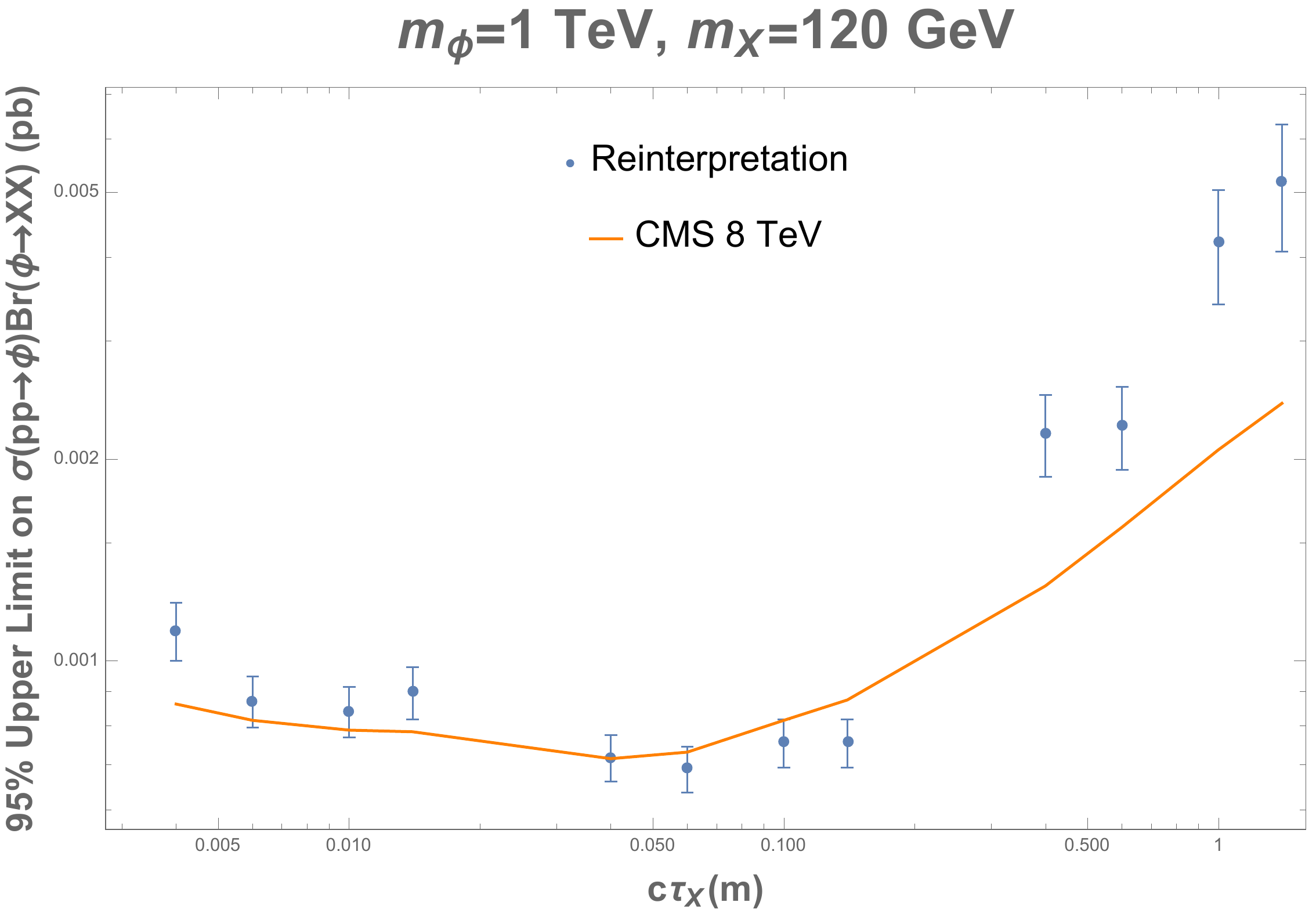}
    \includegraphics[width=0.49\textwidth]{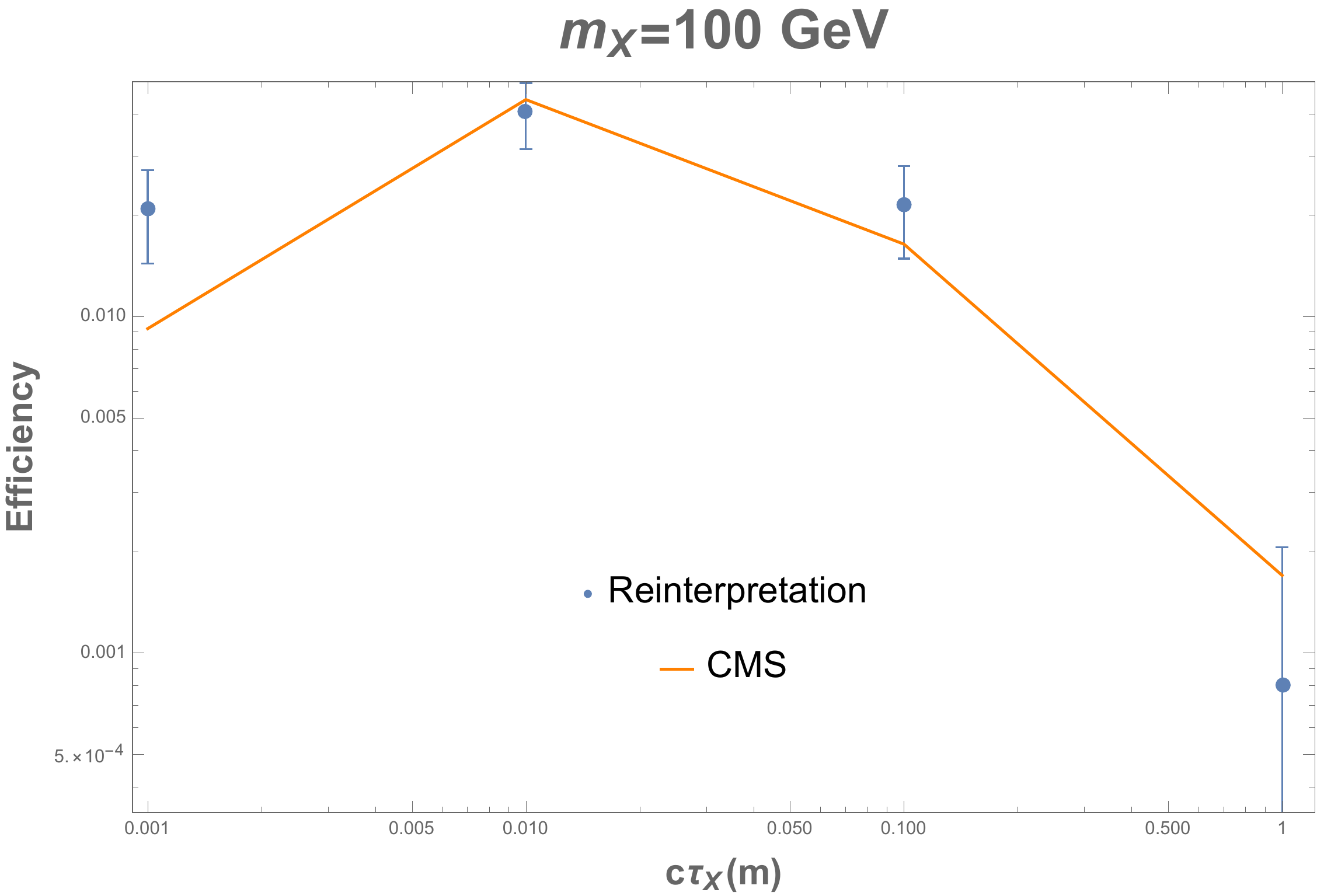}
  \caption{In both panels the {\bf orange line} is the result given by the experimental collaboration and the {\bf blue dots} correspond to the result of our reinterpretation, where the error band are estimated from our montecarlo by taking the variable we are plotting and multiplying it by $2/\sqrt{N}$ where $N$ are the number of events passing the cuts. {\bf Left:} A comparison between the $95\%$ cross section times branching ratio limits placed by the CMS IT analysis at 8 TeV~\cite{CMS:2014wda}  and our reinterpretation as described in Appendix \ref{app:CMS8TeV} for an example mass point, $m_\phi = 1000$ GeV, $m_X = 150$ GeV. {\bf Right:} A comparison between the efficiency reported by the CMS IT analysis  at 13 TeV \cite{Sirunyan:2018vlw} and our reinterpretation as described in Appendix \ref{app:CMS13TeV} for an example mass point $m_X = 100$ GeV. Since no heavy resonance interpretation was directly given by the collaboration we validate our analysis on the jet-jet model of~\cite{Sirunyan:2018vlw} (see text for further details).} \label{fig:cmsITval}
\end{figure}

The CMS IT analysis at 13 TeV \cite{CMS:2014wda} reports a tracker-based search for displaced dijets, and we mirror their event reconstruction and high $\langle L_{xy} \rangle$ selection with the following modifications, some of which are modeled off the reinterpretation performed in \cite{Csaki:2015uza}.  Instead of associating tracks to jets based on angular separation, we reconstruct jets using an anti-$k_T$ algorithm with $R = 0.5$ and use \texttt{Delphes}' assignment of tracks as jet constituents (but still impose that the tracks have $p_T > 1$ GeV to be considered). For the purpose of vetoing jets with `prompt tracks' we define prompt tracks as those whose vertices are located within $0.5$ mm of the primary vertex, rather than placing a requirement on their impact parameters.

To find displaced vertices, rather than running an `adaptive vertex fitter', we identify candidate secondary vertices using a depth-first `grouping' algorithm run on the tracks associated to each possible pairing of jets in the event. Beginning with a single randomly-chosen track as the `seed' track for our algorithm, we look through all other tracks in the pair of jets and create a `group' of tracks consisting of the seed tracks and any others whose origins are within $1$ mm of the seed track. We then add to that group any tracks whose origins are within $1$ mm of the origin of any tracks already in the group, and repeat this step iteratively until no further tracks can be added. We then randomly choose a new seed track which has not yet been assigned to a group and begin the clustering process anew. We repeat this process until all tracks in the pair of jets have been assigned to clusters. We assign to each group a location, which is the average of the origins of all tracks in the group. Any group which contains tracks from both jets in the pair and whose transverse displacement from the beamline, $L_{xy}$, is greater than $2.4$ mm is considered a candidate secondary vertex.

Then for each candidate secondary vertex in the event we further form clusters using the transverse displacement of the tracks in the cluster from the primary vertex, $L_{xy}^\text{track}$. To compute $L_{xy}^\text{track}$ we find the intersection of the track trajectory with the line passing through the origin in the direction of the dijet momentum, as projected onto the transverse plane. We then form a cluster of maximal track multiplicity from tracks associated with the secondary vertex by clustering together tracks whose $L_{xy}^\text{track}$ differ by no more than $0.15 L_{xy}$. We require that this cluster has at least one track from each jet in the dijet, that the cluster has an invariant mass of $> 4$ GeV, and that the vector sum of momenta of tracks in the cluster satisfies $p_T > 8$ GeV. If the event contains more than one secondary vertex which contains clusters satisfying these requirements, we select the secondary vertex of maximal track multiplicity. Finally, the CMS analysis forms a background discriminant using the track multiplicities of the selected vertex and its associated cluster, the RMS of $L_{xy}^\text{track}$ in the cluster, and the signed impact parameter of the tracks in the vertex, and tunes a cut based on the value of this discriminant to give an expected background of $\sim 1$ event in the integrated luminosity considered. We mock up this discriminant by vetoing events where the selected cluster contains $<9$ tracks, which the cluster track multiplicity distribution given in the supplementary data from \cite{CMS:2014wda} shows removes all but $\sim 1$ background event.

This procedure faithfully reproduces the acceptance times efficiency for the heavy scalar simplified model benchmarks presented in \cite{CMS:2014wda}. However, we have found that jet clustering as implemented by \texttt{Delphes} typically forms jets containing significant numbers of particles originating from different secondary vertices. This becomes problematic when extrapolating away from the benchmark signal parameters in \cite{CMS:2014wda} towards signal parameters with larger separation between $m_\phi$ and $m_X$. In this regime, the decay products of the $X$ particles become collimated, such that jets faithfully constructed from the decay products of a single secondary vertex would fail to be resolved and fall out of the acceptance of the search. The tendency for \texttt{Delphes} to cluster particles from different secondary vertices into the same jet leads to an unrealistically high acceptance for signal points with $m_X / m_\phi \ll 1$, as the jet pair passing cuts does not genuinely originate from one vertex. In order to compensate for this inaccurate aspect of our simulation, we additionally require that signal events passing cuts contain at least one secondary vertex whose daughter partons have an angular separation $R \geq 0.5$ at truth level. This has no impact on our reproduction of the efficiencies for the benchmark signal points in \cite{CMS:2014wda}, but leads to an expected falloff in efficiency as $m_X / m_\phi \ll 1$.

For this reinterpretation we generate 5k $pp \rightarrow \phi \rightarrow XX (\rightarrow b\bar b b \bar b)$ signal events at each $(m_X, m_\phi, c\tau_X)$ signal point. \texttt{ROOT} \cite{BRUN199781} is used to perform detailed analysis on reconstruction-level events. We validate our anlaysis by comparing our cross section $\times$ branching ratio limits to benchmarks given in the CMS analysis \cite{CMS:2014wda}, an example of which can be seen in the left panel of Figure \ref{fig:cmsITval}.

 \subsection{CMS Inner Tracker Analysis at 13 TeV}\label{app:CMS13TeV}

The CMS IT analysis at 13 TeV \cite{Sirunyan:2018vlw} updates the previous CMS tracker-based search \cite{CMS:2014wda}. However, the 13 TeV analysis does not require that the reconstructed secondary vertex contains tracks from both of the pair of jets identified as the dijet. As a result it retains sensitivity to the case where pair-produced LLPs each decay to a pair of jets which are collimated and cannot be resolved, but requires more stringent background discrimination. We mock up the event reconstruction and selection described in \cite{Sirunyan:2018vlw} along with the following modifications.

At trigger level we apply a track reconstruction efficiency based on Figure 12 of \cite{Chatrchyan:2014fea}. We ignore tracks with probability given by a linear approximation to the fourth iteration of track reconstruction as a function of the radial distance from the beampipe at which a given track begins. We use the same modified definition of `prompt tracks' as for the 8 TeV reinterpretation above. The definition of `displaced tracks' requires a transverse impact parameter significance larger than $5.0$, so we model a impact parameter uncertainty using a piecewise linear approximation to Figure 20 of \cite{Chatrchyan:2014fea}. We tune the precise values of both of these approximations based on the validation data. Endcap tracks are not treated differently.

At the level of event selection, we go back to truth-level and then apply a track reconstruction efficiency now based on the final iteration of track reconstruction from Figure 12 of \cite{Chatrchyan:2014fea}. We again use our depth-first algorithm to search for secondary vertices, and ignore the $\chi^2$ requirement. We also ignore the cut on the track in the vertex with second-highest transverse impact parameter significance. The analysis also requires the 3D impact parameter resolution, which we again mock up by approximating Figure 20 of \cite{Chatrchyan:2014fea}. We form clusters as above using $L_{xy}^{\text{track}}$ and mock up the discriminant by requiring that the number of tracks in the cluster is greater than 5. We use slightly different definitions of `prompt' when rejecting jets with too many prompt tracks: we require that the jet has no more than one track with 3D impact parameter below $0.3$ mm, and require that tracks with transverse impact parameter below $0.5$ mm comprise no more than $15\%$ of the energy of the jet. 

For this reinterpretation we generate 5k $pp \rightarrow \phi \rightarrow XX (\rightarrow b\bar b b \bar b)$ signal events at each $(m_X, m_\phi, c\tau_X)$ signal point. \texttt{ROOT} \cite{BRUN199781} is used to perform detailed analysis on reconstruction-level events. Since no heavy resonance interpretation was directly given by the collaboration, we validate our analysis by emulating the jet-jet model given in the CMS analysis \cite{Sirunyan:2018vlw}, generating events in which the $X$s are pair-produced through a derivative coupling to the $Z$ boson. We compare the resulting efficiencies to benchmarks given in the CMS analysis, an example of which can be seen in the right panel of Figure \ref{fig:cmsITval}. By tuning our implementation of the track reconstruction probability, the impact parameter resolution, and the discriminant we are able to achieve reasonable agreement. However it is clear that a more faithful reinterpretation will require additional details on the detector response that we do not have access to.

\subsection{CMS 13 TeV `Beampipe' Analysis} \label{app:CMSbp}

The CMS analysis \cite{Sirunyan:2018pwn} reports a search for dijets displaced from the beamline by $0.1$ mm to $20$ mm. A `recipe' is provided for the reinterpretation of the results using truth-level simulated data which was tested to be accurate to $20\%$ for a variety of models with signal efficiency $> 10\%$. We implement this recipe as instructed, with the caveat that we avoid relying on the reconstruction of displaced jet objects. We instead impose an angular separation requirement $\Delta R > 0.4$ on the LLP decay products directly to mimic the requirement that two jets are reconstructed from each decay with the anti-$k_T$ algorithm with distance parameter $0.4$, and place the $H_T$ requirement on the LLP decay products themselves. We generate 5k $pp \rightarrow \phi \rightarrow XX$ signal events at each $(m_X, m_\phi, c\tau_X)$ signal point for this analysis, and give $X$ a branching ratio of $0.8$ to decay to $b\bar b$ and $0.05$ to each of $c\bar c, s \bar s, d \bar d, u \bar u$.

\section{The dark sector of the Fraternal Twin Higgs}\label{app:fraternal}
Here we describe the nature of the dark sector we used as a benchmark for our discussion in Sec.~\ref{sec:twinhiggs} and the approximation we used to estimate the signal rate for displaced events. We focus on the Fraternal Twin Higgs proposal first discussed in Ref.~\cite{Craig:2015pha}. The logic of this bottom up construction is to introduce below a given UV threshold scale $\Lambda_{UV}\gtrsim 5\text{ TeV}$ only the \emph{minimal} amount of states in the dark sector to preserve the naturalness of the Twin Higgs construction. Not surprisingly, this logic fixes the masses and the couplings of only a handful of dark sector states which are the mirror partners of the SM states which couple the most to the SM Higgs. 

The dark top, $t_B$, and the dark gauge bosons, $W_B^i$, are required and demanding naturalness fixes the dark top yukawa $y_{t}^B$ and the dark EW coupling $g_{2}^B$ equal to the SM ones at $\Lambda_{UV}$ within $1\%$ and $10\%$ respectively. A dark bottom, $b_B$, and the dark tau, $\tau_B$, are required for anomaly cancellation in the dark sector, but their coupling can be vastly different from the SM one as long as $y_{b,\tau B}\lesssim y_{tB}$. We assume a dark photon to not be present, as its presence would introduce another portal and modify the phenomenology (see e.g.~Ref.~\cite{Prilepina:2016rlq}). Finally, a dark QCD is also required with a coupling $g_{s}^B$ not far from the SM one.  Given these premises, in what follows we review the structure of the dark sector in the Fraternal Twin Higgs and show the portion of parameter space which is relevant for the phenomenology discussed in this paper.

The Twin sector contains a copy of QCD whose coupling cannot be arbitrarily different from Standard Model QCD because of the approximate mirror symmetry between the two sectors. This leads to dark confinement at a comparable scale to $\Lambda_{\text{QCD}}$. The spectrum of bound states arising from Twin QCD confinement depends sensitively on the fermion spectrum of the dark sector. In the Fraternal Twin Higgs, the Twin QCD near the confinement scale is an $SU(3)$ gauge field theory, with at most one flavor and the lightest states in the confined twin sector being either the glueballs or the bottomonia.

First we identify the Twin QCD confinement scale. Recall that\footnote{Notice that Eq.~35 of \cite{Craig:2015pha} slightly differs with this definition in order to match with the standard lattice definitions \cite{Gockeler:2005rv}.} 
\begin{equation}
\Lambda_{\text{QCD}}=\mu \exp\left(-\frac{1}{2b_0g^2_s}\right)\left(\frac{b_1}{b_0}g_s^2\right)^{-\frac{b_1}{2b_0^2}}\left(1+\frac{b_1}{b_0}g_s^2\right)^{\frac{b_1}{2b_0^2}},\label{eq:QCDconfNLO}
\end{equation}
where
\begin{align}
&b_0=\frac{11}{3} N_c-\frac{2}{3} N_f\ ,\\
&b_1=\frac{34}{3} N_c^2 - \frac{N_c^2 - 1}{N_c} N_f - \frac{10}{3} N_c N_f\ .
\end{align}

A way to define the Twin QCD confinement scale, $\Lambda_{\text{QCD}}^B$, is to use for the renormalization scale, $\mu$, the mass of the lightest \emph{active} flavor in the theory. We first define the dark strong coupling constant $\alpha_s^{B}\equiv(g_s^{B})^2/4\pi$ by imposing a boundary condition at the UV cut-off, $\Lambda$, where we expect the $\mathbb{Z}_2$-symmetry relating the Standard Model QCD and its counterpart to be satisfied up to small deviations,
\begin{equation}
g_s^{B}(\Lambda)=g_s^A(\Lambda)+\delta g.\label{eq:cutoffchange}
\end{equation}
The running of $\alpha_s^{B}$ is then the one of a gauge theory with $N_c=3$ but at most 2 active flavors, the Twin top and the Twin bottom. As a consequence, the Twin strong coupling constant will run faster in the IR. We also allow the dark bottom Yukawa to be modified such that 
\begin{equation}
m_{b_B}= m_b\cdot \frac{f}{v}\cdot \delta y_b,
\end{equation}
where $m_b$ is the SM bottom quark mass computed at the $\Lambda_{\text{QCD}}^{B}$ scale. If $m_{b_B}>\Lambda_{\text{QCD}}^{B}(m_{t_B})$ then we take $\Lambda_{\text{QCD}}^{B}(m_{b_B})$ in Eq.~\eqref{eq:QCDconfNLO} with $N_f=0$ while, in the other limit, we use $\Lambda_{\text{QCD}}^{B}(m_{t_B})$ in Eq.~\eqref{eq:QCDconfNLO} with $N_f=1$.

\begin{figure}[t]
\centering
\includegraphics[width=.48\textwidth]{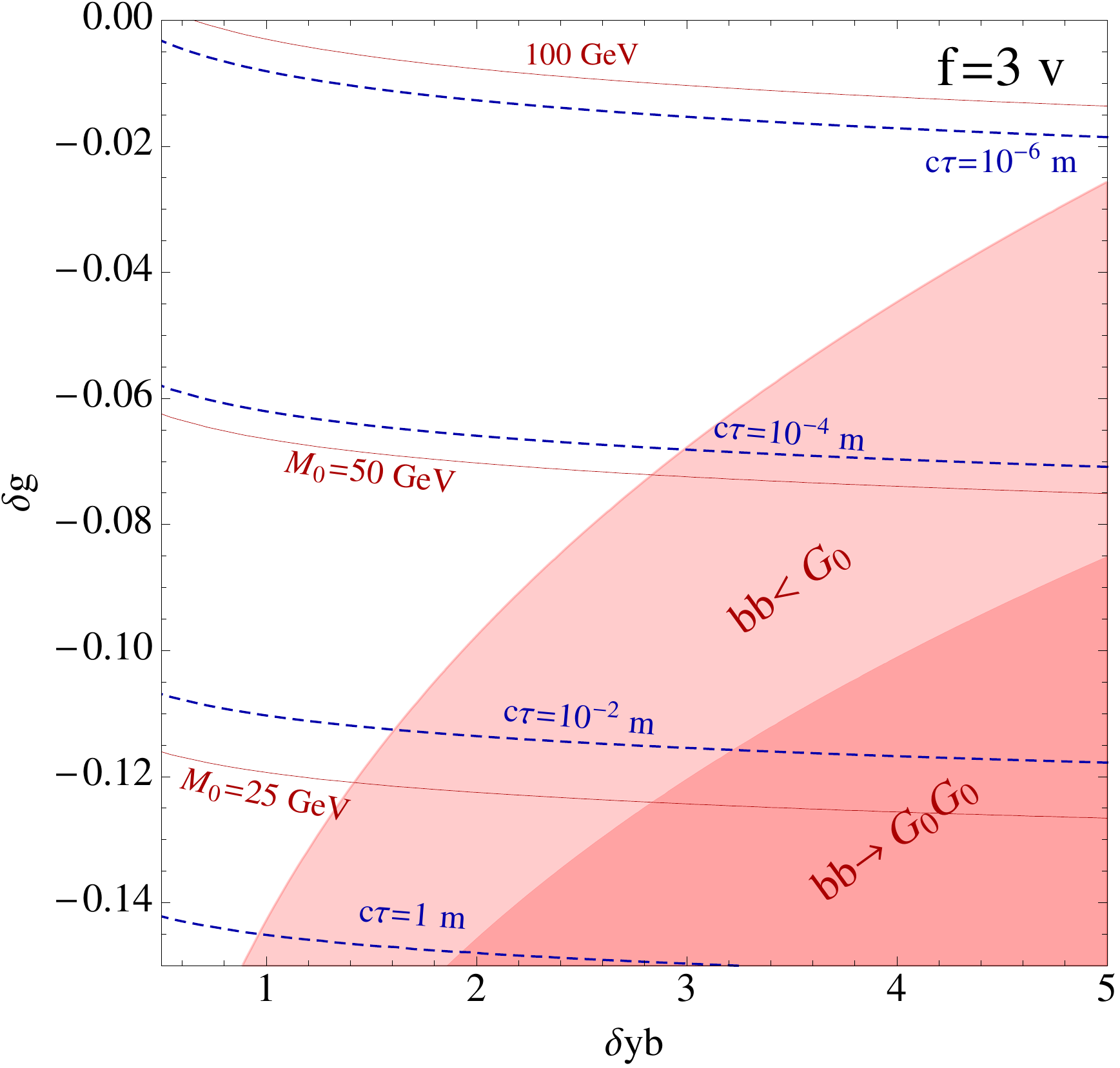}\hfill
\includegraphics[width=.48\textwidth]{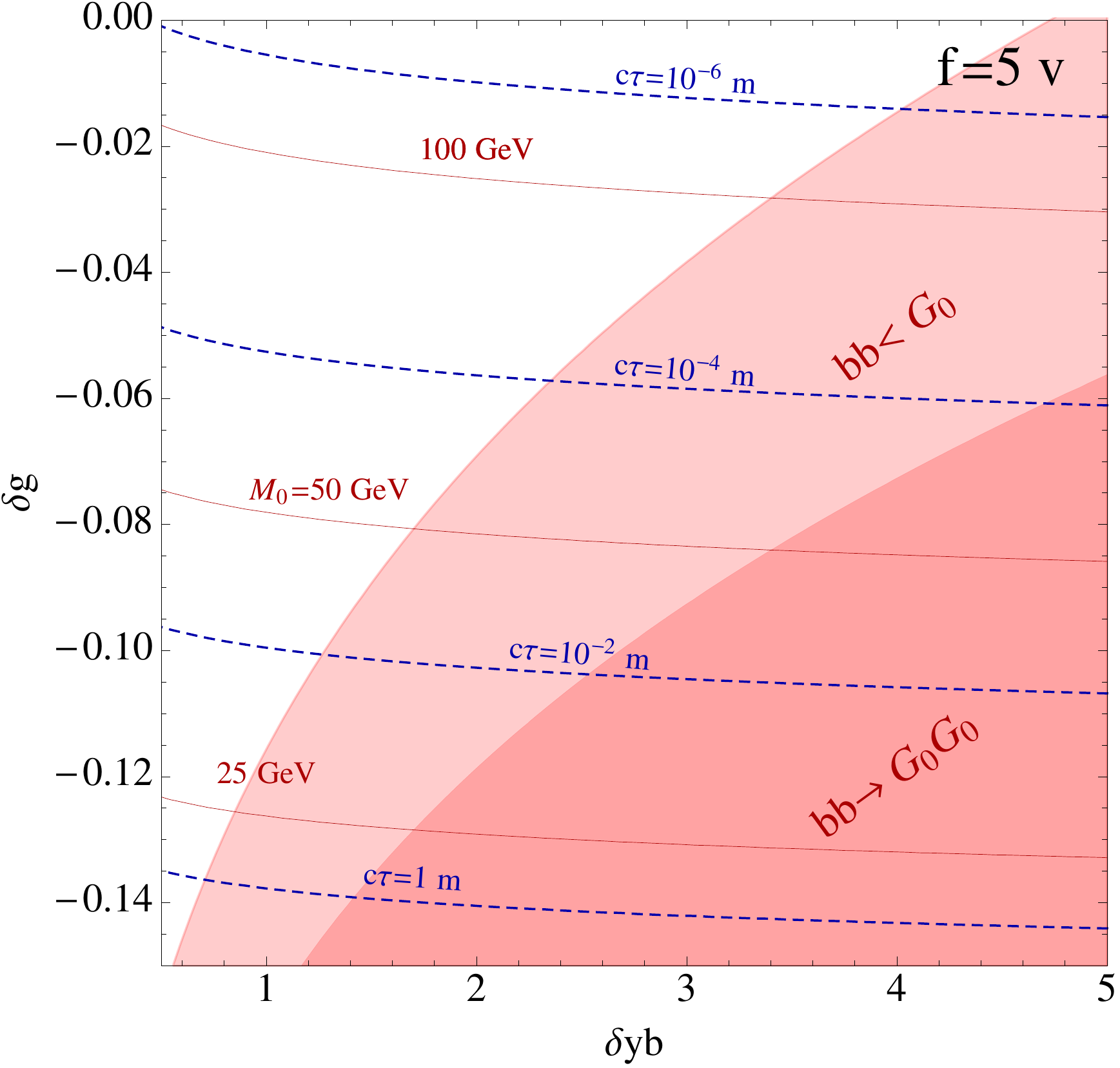}
\caption{Summary of the structure of the dark sector in the Fraternal Twin Higgs. The light pink shaded region shows where the $0^{++}$ glueball, i.e. $G_0$, is the LDP, darker shading indicates where the decay of bottomonia into glueballs pairs is allowed. Red contours indicate the mass of the LDP. Blue dashed contours denote the average $c \tau$ of $G_0$. {\bf Left:} We fix the Twin Higgs VEV to $f=3v$.  {\bf Right:} We fix the Twin Higgs VEV to $f=5v$.
\label{fig:DarkQCD}
}
\end{figure}

Given $\Lambda_{\text{QCD}}^B$, we can develop a more quantitative picture of the Twin spectrum. The lightest glueball carries $J^{PC}$ quantum numbers $0^{++}$, and its mass is related to the Twin QCD scale as \cite{Gockeler:2005rv,Morningstar:1999rf}
\begin{equation}
M_0=m_{G_{0}^{++}}=6.8\Lambda_{\text{QCD}}^B.
\end{equation}
The masses of the additional stable glueball states are set in terms of $M_0$ by lattice computations on $N_f=0$ QCD \cite{Gockeler:2005rv}\footnote{The two lattice computations agree very well but for the presence of the extra excited $0^{++}$ state.}:
\begin{align}
& m_{G_{2}}^{++}=1.4 M_0, ~~m_{G_{0,\,(2)}^{++}}=1.54 M_0,\\
& m_{G_{0}^{-+}}=1.5 M_0, ~~m_{G_{2}^{-+}}=1.8 M_0,\\
& m_{G_{1}^{+-}}=1.7 M_0,~~ \dots
\end{align}
 For the spectrum of the stable bottomonia, we take  \text{Ref.~\cite{Patrignani:2012an}}
\begin{equation}
 m_{\chi_{0}^{-+}}\lesssim m_{\chi_{1}^{--}}\lesssim m_{\chi_{1}^{+-}}\lesssim m_{\chi_{0}^{++}} \lesssim m_{\chi_{1}^{++}}=2 (m_{b_B}+\Lambda_{\rm{QCD}}^B).
 \end{equation}
Depending on $\delta g$ and $\delta y_b$ the Lightest Dark Particle (LDP) can be either the CP-odd bottomonium or the CP even glueball. We will focus here in the case where $M_0\lesssim  m_{\chi_{0}^{-+}}$, as highlighted by the spectrum presented in Fig.~\ref{fig:DarkQCD}. As one can see, in the most natural region of the Twin Higgs parameter space (i.e. $f\lesssim 5v$), $M_0$ is not allowed to be heavier than roughly 100 GeV. 
 
\paragraph{Dark hadronization and spectrum} 

\paragraph{Twin-SM portals} 
There are two leading portals that connect the bound states of Twin QCD confinement to the Standard Model. At low energies, below the mass of the Twin Higgs, they can be written in terms of higher-dimensional operators that mix the $B$ sector with the $A$ sector: 
\begin{equation}\label{eq:HiggsPortal}
\mathcal{L}^{\text{H-portal}}_{\text{SM}}= \frac{c_{H}^{gg} H_A H_A^\dagger}{\Lambda^2}G_{B}\cdot G_B+\frac{c_{H}^{bb} H_A H_A^\dagger}{\Lambda} \bar{b}_Bb_B\ .
\end{equation}
These portal operators both allow the SM-like Higgs to decay into two or more Twin QCD bound states and enable the bound states with appropriate quantum numbers to mix with the Higgs. While there are additional portal operators beyond these, they appear at higher dimension or involve states neutral under Twin QCD.

The Wilson coefficients of both operators have irreducible infrared contributions of the form
\begin{align}
&\frac{c_{H}^{gg}}{\Lambda^2}=\frac{\alpha_s^B}{12\pi f^2}+\dots\,,\label{eq:gluglu}\\
&\frac{c_{H}^{bb}}{\Lambda}= \frac{y_b}{f}+\dots\label{eq:bb}
\end{align} 
The contribution in Eq.~\eqref{eq:gluglu} comes from integrating out the Twin tops in the infinite mass limit, while the contribution in Eq.~\eqref{eq:bb} arise from the Twin bottom Yukawa coupling after mixing is taken into account. Thanks to the first operator in (\ref{eq:HiggsPortal}), the scalar lightest glueball $0^{++}$ will be unstable and will decay back to the SM in a Higgs-like manner, with a width suppressed by $v^2/\Lambda^2$. The width of the $0^{++}$ glueballs is given by

\begin{equation}
\Gamma (G_0)=\Gamma_h (m_0)\left(2c_3\frac{v}{\Lambda^2}\frac{\kappa m_0^3}{m_h^2-m_0^2}\right)^2,
\end{equation}
where $\kappa$ parametrizes the matrix element $\langle 0|G_B\cdot G_B|G_0\rangle =F_0^3=\kappa m_0^3$, that can be extracted from the lattice data ($\kappa\simeq 0.25$), while $\Gamma_h (m_0)$ is the width of a Standard Model Higgs boson with mass $m_0$. In Fig. \ref{fig:DarkQCD}, we show in blue the lightest glueball proper life time as a function of $\delta y_b$ and $\delta g$; the red curves represent the glueball mass in GeV. For the plots, we fix $c_3=1$ and $\Lambda=2\pi f=6\pi v$. Correspondingly, also the heavier $0^{++}$ glueballs will have Higgs-like decays thanks to its mixing with the Higgs boson with macroscopic decay lengths. 

The glueball states with other $J^{PC}$ quantum numbers are either stable or quasi-stable, depending on the detailed spectrum in the Twin sector and the available decay modes. Even when decay modes invariably exist, such as the decay of the $2^{++}$ state via an off-shell Higgs boson ($2^{++}\to 0^{++}h^*$), the resulting lifetimes are typically too long to give appreciable decays inside the LHC detectors, and instead generically lead to missing energy. In its totality, the pure glueball spectrum contains a variety of states, of which two decay back to the SM with Higgs-like branching ratios and with a macroscopic life-time \cite{Juknevich:2009gg}.

\section{Final states and rates in the Twin Higgs}\label{sec:ratesTwin}

\begin{figure}[th!]
\centering
\includegraphics[width=14cm]{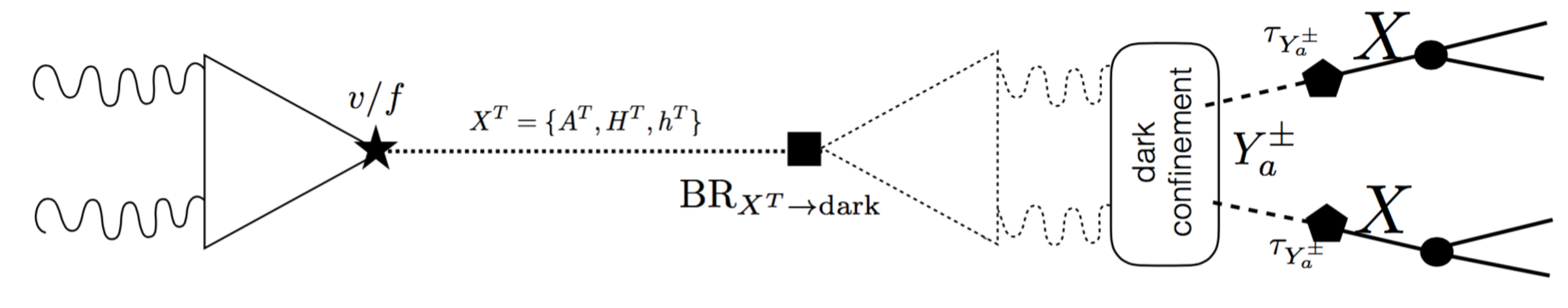}
\caption{Cartoon of the production and decay of LLPs through heavy Twin Higgs states.
\label{fig:cartoon_1}
}
\end{figure}

In Fig.~\ref{fig:cartoon_1} we sketch the structure of the displaced processes of interest. Since the process is fairly complicated, it is useful to factorize it into different building blocks that we discuss separately. 

The gluon fusion cross section for a scalar $\Phi$ can be written at leading order as 
\begin{equation}
\sigma_{LO}(gg\to \Phi)=\frac{\pi^2}{8 m_\Phi^3}\Gamma_{LO}(\Phi\to g g)\ .
\end{equation}
where $\Gamma_{LO}(\Phi\to g g)$ is the scalar width into gluons (and is given in Eq.~\eqref{eq:SMgluon} for the Twin Higgs). This cross section is known to be enhanced at NLO by 60\%-90\%, depending on the mass of the scalar. 
For our numerical calculations, we take the state-of-the-art NNLO+NNLL values for the SM Higgs cross section from \cite{MelladoGarcia:2150771}. The Twin Higgs cross section at $\sqrt s$ will be given by 
\begin{equation}
\sigma_{gg\to \phi}(m_\phi,\sqrt{s})=\sin^2\theta \sigma^{{\text{SM}}}_{gg\to h}(m_\phi,\sqrt{s}).
\end{equation}

The partial widths of the Twin Higgs to dark sector particles are
\begin{align}
&\Gamma( \phi\to t_B t_B)=\frac{3 m_t^2 m_{ \phi}}{16\pi v^2}\cos\theta^2\left(1-4\frac{m_{t_B}^2}{m_{ \phi}^2}\right)^{3/2}\ ,\\
&\Gamma( \phi\to b_B b_B)=\frac{3\delta y_b^2 m_b^2 m_{ \phi}}{16\pi v^2}\cos\theta^2\left(1-4\frac{m_{b_B}^2}{m_{ \phi}^2}\right)^{3/2}\ ,\\
&\Gamma( \phi\to V_B V_B)= s_V \frac{g^2_B m_{ \phi}^3}{64\pi M_{V_B}^2}\cos\theta^2\left(1-4 \frac{m^2_{V_B}}{m^2_{ \phi}}\right)^{1/2}\left(1-4\frac{m^2_{V_B}}{m^2_{ \phi}}+12\frac{m^4_{V_B}}{m^4_{ \phi}}\right)\ ,\\
&\Gamma( \phi\to g_B g_B)=\frac{\alpha_{s B}^{2} m_{ \phi}^3}{144\pi^3 f^2}\cos\theta^2\left\vert \sum_{Q=t,b} A\left(\frac{4 m_Q}{m_{ \phi}}\right)\right\vert^2\ ,\quad A(\tau)\equiv\frac{3}{2}\tau(1+(1-\tau) f(\tau))\,,
\end{align}
where $s_V=1\, , (1/2c_W^2)$ for $V=(W, Z)$ and $g_B$ is the Twin gluon. $f(\tau)$ is the standard gluon fusion loop function.
The LO decays into SM objects are\footnote{Notice that the width in Eq.~\eqref{eq:SMhiggs} has a factor of 1/2 with respect to B.29 of \cite{Katz:2016wtw} which is due to the different definition of $A_{ \phi hh}$. }
\begin{align}
&\Gamma( \phi\to f_A f_A)=\frac{3 m_f^2 m_{ \phi}}{16\pi v^2}\sin\theta^2\left(1-4\frac{m_{f}^2}{m_{ \phi}^2}\right)^{3/2}\ ,\\
&\Gamma( \phi\to V_A V_A)=s_V \frac{g^2 m_{ \phi}^3}{64\pi M_{V_A}^2}\sin\theta^2\left(1-4 \frac{m^2_{V_A}}{m^2_{ \phi}}\right)^{1/2}\left(1-4\frac{m^2_{V_A}}{m^2_{ \phi}}+12\frac{m^4_{V_A}}{m^4_{ \phi}}\right)\ ,\\
&\Gamma( \phi\to h h)= \frac{A_{ \phi hh}^2}{32\pi m_{ \phi}}\left(1-4\frac{m_h^2}{m_{ \phi}^2}\right)^{1/2}\ ,\label{eq:SMhiggs}\\
&\Gamma( \phi\to g_A g_A)=\frac{\alpha_{s A}^{2} m_{ \phi}^3}{144\pi^3 f^2}\sin\theta^2\left\vert \sum_{Q=t,b} A\left(\frac{4 m_Q}{m_{ \phi}}\right)\right\vert^2\,,\label{eq:SMgluon}
\end{align}
where we have defined the triplet interaction $A_{ \phi hh}$ in (\ref{eq:trilinearhT}). NLO corrections can shift the decay into Twin gluons by $\sim50\%$ \cite{Spira:2016ztx}. These can be included following \cite{Spira:1995rr}. As expected, in the limit $\lambda\gg g^2\, ,\kappa\, ,\rho$ (i.e. for large mass of the Twin Higgs) we recover the result from the Goldstone equivalence theorem
\begin{align}
&\Gamma( \phi\to Z_{A,B} Z_{A,B})\approx \frac{1}{2}\Gamma( \phi\to W_{A,B} W_{A,B})\approx \Gamma( \phi\to h h)=\frac{\lambda m_{ \phi}}{8\pi}\,,\\
&\text{BR}( \phi\to Z_{A,B} Z_{A,B})\approx \frac{1}{2}\text{BR}( \phi\to W_{A,B} W_{A,B})\approx \text{BR}( \phi\to h h)=\frac{1}{7}\,,
\end{align}
where the branching ratios in the different channels approach the same constant. In this limit we can also estimate the narrow width limit for the Twin Higgs resonance 
\begin{equation}
\Gamma/m_{ \phi}\lesssim 1\quad \Rightarrow  \quad \lambda\lesssim \pi.
\end{equation} 
The resulting upper bound on $\lambda$ still allows a large portion of the parameter space of the Composite Twin Higgs to be probed by LHC narrow width resonance searches.  

 One important lesson we want to emphasize is that for $\rho\neq 0$ the mixing receives important corrections (bigger than $10\%$) in the region where the Twin Higgs is lighter than 1 TeV while the trilinear coupling is always well approximated by its leading order expression. Expanding the full expressions, we get
 \begin{align} \label{eq:thangle}
&\sin^2\theta =  \frac{v^2}{f^2} -\frac{m_h^2 }{m_\phi^2-m_h^2}\Big(1-2\frac{v^2}{f^2}\Big) + \frac{2 \rho v^2}{m_\phi^2-m_h^2} \Big(1-\frac{v^2}{f^2}\Big)\,,\\
&A_{ \phi hh} \simeq \frac{m_\phi^2}{f} \left[1- \frac{\lambda - \rho }{2 \lambda} \right].
 \end{align}
This effect makes the branching ratio into SM Higgs pairs dominant in the light mass region for the Twin Higgs somehow weakening the bounds from $ZZ$ decays with respect to Goldstone theorem limit. For $\rho\neq0$ also the trilinear coupling gets modified. 

\begin{figure}[t]
\includegraphics[width=.5\textwidth]{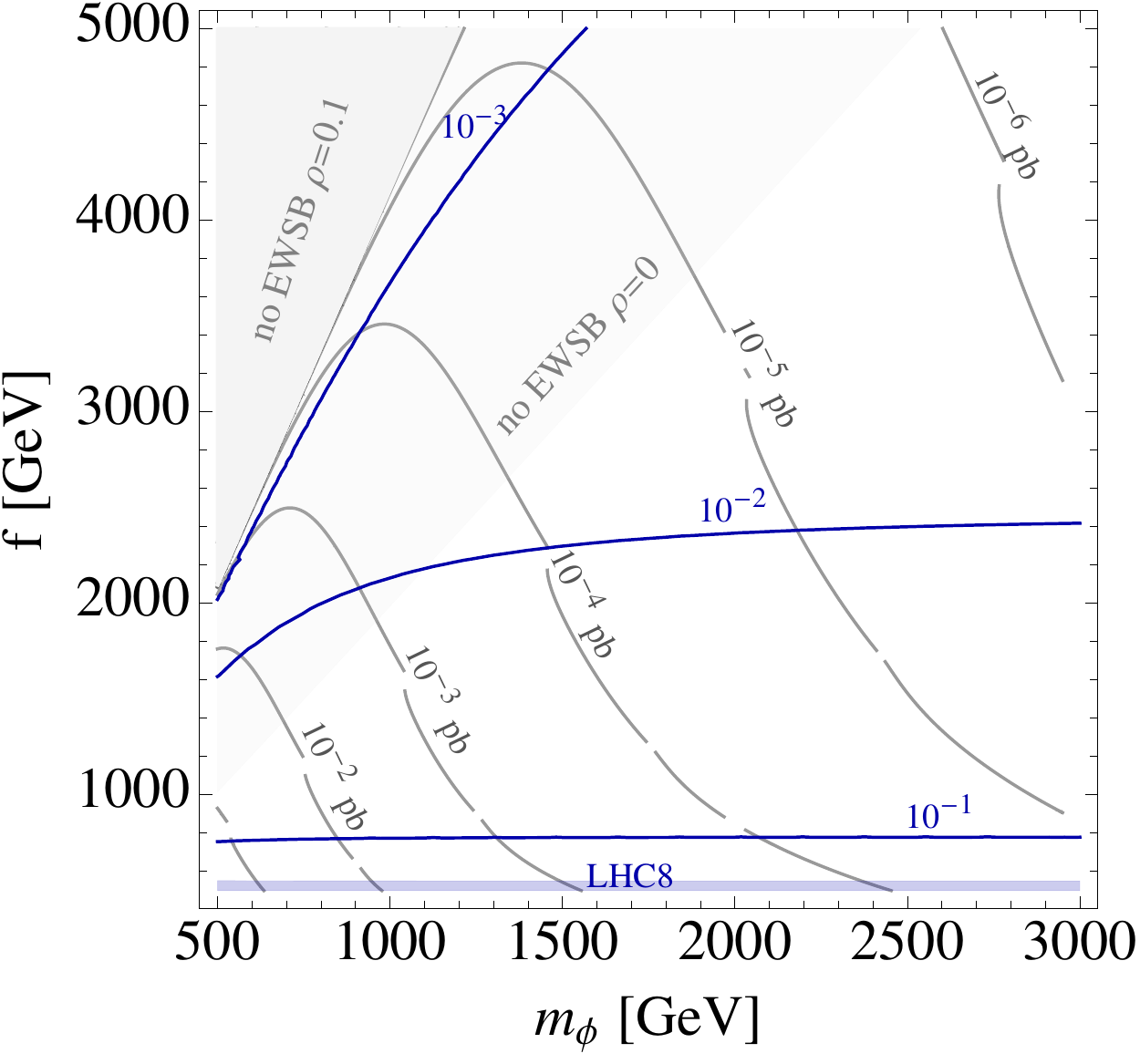}\hfill
\centering
\caption{\label{fig:twinrate} Rates for the Fraternal Twin Higgs production cross section times branching ratio into long-lived glueballs as a function of the Twin Higgs mass, $m_\phi$, and $f$. In (light) gray, we show the region that cannot produce successful EWSB for $\rho=0.1$ ($\rho=0$). The blue shaded region at low values of $f$ is probed by 8 TeV Higgs coupling measurements. The solid blue lines indicate values of $\sin^2 \theta$ using \eqref{eq:thangle}.
}
\end{figure}

The production of glueballs through decays of the heavy Twin Higgs proceeds through a variety of channels. While the Twin Higgs has direct decays into Twin gluons, this rate is relatively small. A far larger production rate arises from decays into heavier dark sector states which undergo subsequent annihilations and/or decays. In particular, pair production of Twin bottoms leads to annihilation decays into Twin glueballs when kinematically accessible. Likewise, pair-produced Twin tops decay into $W_B W_B b_B \bar b_B$ final states, while pair-produced Twin $Z$ bosons decay (in part) to $b_B$ pairs. In both cases these $b_B$ undergo subsequent annihilation decays into Twin glueballs. Of course, the final states resulting from these processes are rich complex, with a variety of both long-lived and detector-stable particles and widely varying multiplicities. The presence of additional detector-stable particles in the final state may reduce the sensitivity of some LLP searches that veto missing energy, while providing stronger coverage from searches that combine missing energy with displaced vertices. Detailed study of the final states necessarily requires a careful treatment of dark showering and hadronization, which is beyond the scope of the present work. For the sake of illustration, we estimate the branching ratio into glueballs coming from the fraction of decays of the Twin Higgs into $t_B \bar t_B$ and $Z_B$ $Z_B$ to be 100\%. The resulting production cross section times branching ratio into glueballs of the heavy Twin Higgs is shown in Fig. \ref{fig:twinrate} for $\rho = 0, 0.1$.
  
\bibliographystyle{JHEP}
\bibliography{biblio_Exotic}

\end{document}